\documentclass[12pt,tightenlines,eqsecnum,floats,aps,amsmath,amssymb,tightenlines,nofootinbib,prd,shownopacs,notitlepage]{revtex4}

\usepackage{graphicx,verbatim}
\usepackage{amsmath}
\usepackage{amsfonts}
\usepackage{amssymb}
\usepackage[colorinlistoftodos]{todonotes}
\usepackage{epstopdf}

\def\rmd{\mathrm{d}}
\def\={\hat{=}}

\newcommand{\pb}[1]{\hbox{\lower0.5ex\hbox{${}_{\leftarrow}$}}\kern-1.9ex{#1}}

\def\t{\tilde}

\def\be{\begin{equation}}
\def\ee{\end{equation}}
\def\ba{\begin{eqnarray}}
\def\ea{\end{eqnarray}}

\def\f{\frac}

\def\Do{\mathring{D}}

\def\vx{\vec{x}}

\def\hphi{\hat{\phi}}
\def\hpi{\hat{\pi}}

\def\hphi{\hat{\Phi}}
\def\hpi{\hat{\Pi}}
\def\ha{\hat{a}}
\def\had{\hat{a}^{\dag}}

\def\ak{\hat{A}_{\vec{k}}}
\def\adk{\hat{A}_{\vec{k}}^{\dagger}}
\def\admk{\hat{A}_{-\vec{k}}^{\dagger}}
\def\adag{\hat{A}^{\dagger}}

\def\hphik{\hat\phi_{\vec{k}}}
\def\vkp{\vec{k}^{\prime}}
\def\vp{\varphi}
\def\hvp{\hat\varphi}
\def\vpk{\varphi_{\vec{k}}}
\def\hvpk{\hat{\varphi}_{\vec{k}}}

\def\hpi{\hat\pi}
\def\hpik{\hat{\pi}_{\vec{k}}}
\def\hpikp{\hat\pi_{\vec{k}^{\prime}}}

\def\vk'{\vec{k}^{\prime}}
\def\vk{\vec{k}}

\def\hx{\hat{x}}
\def\hp{\hat{p}}

\begin{document}

\title{ Emergence of classical behavior in the early universe}

\author{Abhay Ashtekar${}^{1}\,$}
\email{ashtekar.gravity@gmail.com} 
\author{Alejandro Corichi${}^{1,2}\,$}
\email{corichi@matmor.unam.mx}
\author{Aruna Kesavan${}^{1}\,$}
\email{aruna.kesavan@gmail.com}
\affiliation{
${}^{1}$ Institute for Gravitation and the Cosmos \& Physics Department, The Pennsylvania State University, University Park, PA 16802 U.S.A.\\
\\
${}^{2}$ Centro de Ciencias Matem\'aticas, Universidad Nacional Aut\'onoma de 
M\'exico,UNAM-Campus Morelia, A. Postal 61-3, Morelia, Michoac\'an 58090, Mexico
}

\begin{abstract}

{ We investigate three issues that have been discussed  in the context of inflation: Fading of the importance of quantum non-commutativity; the phenomenon of quantum squeezing; and the ability to approximate the quantum state by a distribution function on the classical phase space. In the standard treatments, these features arise from properties of mode functions of quantum fields in (near) de Sitter space-time. Therefore, the three notions are often assumed to be essentially equivalent, representing different facets of the same phenomenon. We analyze them in general Friedmann-Lema\^itre- Robertson-Walker space-times, through the lens of geometric structures on the classical phase space. The analysis shows that: (i) inflation does not play an essential role; classical behavior can emerge  much more generally;  (ii) the three notions are conceptually distinct; {classicality can emerge in one sense but not in another;} and, (iii) the third notion is realized in a surprisingly strong sense; there is \emph{exact} equality between completely general $n$-point functions in the classical theory and those in the quantum theory, provided the quantum operators are Weyl ordered. These features arise already for linear cosmological perturbations by themselves: considerations such as mode-mode coupling, decoherence, and measurement theory --although important in their own right-- are not needed for emergence of classical behavior in any of the three senses discussed. Generality of the results stems from the fact that they can be traced back to geometrical structures on the classical phase space, available in a wide class of systems. Therefore, this approach may also be useful in other contexts.}
\end{abstract}

\maketitle

\section{Introduction}
\label{s1}

Current cosmological theories provide a striking picture of cosmogenesis. The very early universe is extremely well approximated by a spatially homogeneous and isotropic Friedmann, Lema\^{i}tre, Robertson, Walker (FLRW) space-time, together with cosmological perturbations, represented by quantum fields. {Furthermore, the Heisenberg state of these fields is a spatially homogeneous and isotropic `vacuum'.}  However, unlike their classical counterparts,  quantum  fields are subject to uncertainty relations that lead to inevitable quantum fluctuations. As the universe expands, these fluctuations are stretched and lead to the { anisotropies} seen in the cosmic microwave background (CMB), which in turn serve as seeds for formation of the large scale structure. Thus, the origin of the large scale structure of the universe is traced back to quintessential quantum fluctuations that cannot be switched off even in principle.

For concreteness, let us consider the inflationary scenario. Then the quantum fields are assumed to be in the Bunch-Davies vacuum, tailored to the near de Sitter symmetry during (the relevant phase of the) slow roll. But in actual calculations, at the end of inflation one replaces the Bunch-Davies vacuum with a distribution function on the classical phase space and describes the subsequent evolution in classical terms. Therefore a number of natural questions arise.  Why is this procedure justified, given that the quantum nature of fluctuations was essential to begin with? Can one justify this approximation from first principles?  In other words, in what precise sense does the classical behavior emerge even though  the starting point is quintessentially quantum? Is inflation essential for this emergence of classicality? Or, is it a general feature of quantum field theory in expanding cosmologies? Is it essential to make a division of quantum perturbations into the `system'  and `environment'  and use the ideas of decoherence? Is it essential to consider non-linear mode-mode couplings, or, can classical behavior emerge in a precise sense even in the linear approximation for quantum perturbations? Is one forced to bring in considerations from measurement theory and use models of wave function collapse a la, say, \cite{diosi1,grw,diosi2,diosi3,bg,pearle,penrose1,penrose2} or, alternatively, the de Broglie-Bohm version of quantum theory, a la, say \cite{netoetal} in which there is no collapse? Because these issues are conceptually important, there is a  large body of literature that addresses them from a variety of  perspectives (see, e.g., \cite{guthpi,lpg,albretcht,dpaas,lps,kiefer1,kiefer2,psd,decoherence1,decoherence2,decoherence3,decoherence4,decoherence5,decoherence6,decoherence7,decoherence8,decoherence9,ls,decoherence10,jmvv1,sss,jmvv2,jmvv3,jmvv4}). 
Our emphasis will be on isolating the simplest mechanisms that can lead to classical behavior in the early universe. 

More precisely, the goal of this paper is three-fold. \emph{First,} we will show that three of the commonly used notions of emergence of classical behavior are not equivalent;  a quantum system can exhibit classical behavior in one sense but not in another. In particular, classical behavior emerges also in a wide class of non-inflationary backgrounds, including radiation and dust filled universes in two of the three senses, but not in the third.%
\footnote{\label{fn1} While this general feature is significant purely from a conceptual viewpoint, this fact is also physically quite interesting because, in the standard inflationary scenario the universe undergoes even more e-folds in its expansion during the radiation and dust dominated eras than during the relevant slow roll phase of inflation, i.e. the epoch between the time when the mode with the largest observable wavelength exits the Hubble horizon till the end of inflation when the slow roll parameter $\epsilon$ becomes $1$.}
Our discussion will emphasize novel features that emerge as the system evolves,  illuminate the underlying mechanisms, and, in some cases,  correct inaccuracies in the literature. \emph{Second,} the discussion will also show that  classical behavior already appears in the mathematical theory of linear perturbations. Considerations such as decoherence and measurement theory in quantum mechanics are, of course, important and are likely to play important roles in a complete understanding of dynamics. But classical behavior emerges without them in the early universe. \emph{Third,} we will highlight certain \emph{geometrical structures} on the \emph{classical phase space} that play an important role in sharpening the sense in which classical behavior emerges. In particular, we will show that the origin of quantum squeezing can be traced back to these geometrical structures. They also provide a natural avenue to associate with the quantum vacuum a distribution function on the classical phase space that, in turn, leads to a stronger result on the relation between quantum and classical $n-$point functions, including those that feature both field operators \emph{and} their conjugate momenta.
 
The paper is organized as follows.  In Section \ref{s2}, we collect a few facts about quantum fields in FLRW space-times, emphasizing the key mathematical input that is needed in the passage from the classical to the quantum theory  --introduction of a K\"ahler structure on the phase space, compatible with the symplectic structure thereon.  Basic concepts and the notation introduced in this discussion will be used throughout the rest of the paper. The next three sections are devoted to the three different notions of emergence of classical behavior. 

One of the first arguments for this emergence in the context of inflation was that quantum `non-commutativity becomes negligible because one can ignore the decaying mode'  (see, e.g. \cite{dpaas,kiefer1,kiefer2}).  In Section \ref{s3} we re-examine this idea in light of the fact that the commutator between the field and its canonically conjugate momentum is constant throughout the evolution and thus cannot become negligible under time evolution. 
We will show that there is indeed  a precise sense in which non-commutativity `fades' during inflation, but it is more subtle.  Appendix \ref{a1} revisits the issue of fading of non-commutativity discussed in section \ref{s3}, but now for the commutators between field operators at different times. 

In Section \ref{s4} we discuss quantum squeezing, which is often taken to be another hallmark of the emergence of classical behavior (see, e.g., \cite{guthpi,lpg,albretcht,dpaas,lps,kiefer1,kiefer2,jmvv1}).  We show that the origin of this phenomenon can be directly traced back to the K\"ahler metric on the phase space. It appears that this `geometrical underpinning' of  squeezing has not been noticed before, at least in the cosmological context.  It serves to bring out the fact that although squeezing is discussed almost entirely in the context of inflation in the cosmology literature, it occurs much more generally during the cosmic expansion, in particular in radiation and dust filled universes where squeezing is in fact more extreme in a precise sense. Interestingly, in this case, classical behavior does \emph{not} emerge in the sense of section \ref{s3}; thus the notions are inequivalent!

In section \ref{s5} we show that the K\"ahler geometry considerations of Section \ref{s4} also provide a natural avenue to associate a phase space distribution function $\rho_{o}$ with every spatially homogeneous, isotropic quantum vacuum $\Psi_{o}$ of the field, on \emph{any} FLRW background. We then show that the evolution of the expectation values in the state $\Psi_{o}$ of any finite product of field operators \emph{and} their conjugate momenta is \emph{exactly} reproduced in the expectation values in $\rho_{o}$ of the product of the corresponding (commuting) classical observables, provided one uses the Weyl (i.e.  totally symmetric) ordering of quantum operators. Thus, the difference between classical and quantum evolutions arises only if we have an operator product that has \emph{both} field operators and their momenta, and the product is \emph{not} Weyl ordered.  This statement provides a complete characterization of the precise difference between classical and quantum predictions (to the extent that both theories are determined by the expectation values of all these $n$-point functions). Results of Section \ref{s5} admit a direct generalization to linear quantum fields in any globally hyperbolic space-time. 

In Section \ref{s6} we summarize the main results emphasizing new elements. As is clear from the above discussion, our focus is on clarifying the sense in which the dynamics of quantum fields representing cosmological perturbations can exhibit behavior that we normally associate with classical systems. In this discussion, then, there is no `quantum to classical transition'. Hence we will not need to enter a discussion of issues that arise when the focus is on this  `transition': decoherence,  quantum measurement theory, collapse of the wave function,  or reformulations of quantum mechanics, e.g., a la de Broglie and Bohm in which there is no collapse.  Rather, our emphasis is on the `emergence' of classical behavior in the early universe in the mathematical description of cosmological perturbations. 

It is then natural to ask: Can the emergence of classical behavior be discussed without entering into the details of the quantum measurement theory?  To conclude this section,  we will make a 
brief detour into \emph{imperfect measurements} to illustrate why this is possible already for familiar \emph{macroscopic} systems. Consider, as a simple example, the pendulum in a grandfather clock. Suppose the mass of the pendulum is $1$kg and frequency  of oscillations is  $\omega= 1\,{\rm s}^{-1}$.  Now suppose the pendulum is in its ground state and we measure its position.  In the Copenhagen interpretation, a perfect measurement will collapse the wave function drastically, giving the pendulum infinite momentum and destroying the clock; the quantum behavior will be very different from the classical prediction! But this is not what one does in practice. When we observe the pendulum, we measure the position to a very good but finite accuracy, say of $\epsilon_{x} = 10^{-5}$m.  And indeed we can make this imperfect measurement repeatedly. Each of these measurements disturbs the pendulum. The momentum imparted is $\hbar/\epsilon_{x} = 10^{-29}$kg m/s whence the velocity is  $10^{-29}$m/s.  We can detect this velocity through a change in its position. But for the displacement to be measurable, i.e. $\sim 10^{-5}m$, we would have wait some $9.5\times10^{23}$s, or $\sim 3 \times 10^{16}$years, or  $\sim 2\times 10^{6}$ times the age of the universe! Thus,  if one makes an imperfect measurement on a macroscopic system --and all our measurements of the position of the pendulum are imperfect--  the system \emph{is} disturbed because of the measurement. However,  even in the  standard Copenhagen interpretation  --even without other considerations such as environment, decoherence, or the de Broglie Bohm version of quantum mechanics-- the resulting collapse of the wave function does not affect the future evolution in any significant manner, provided the window $\epsilon_{x}$ of the imperfect measurement is much larger than the Heisenberg  uncertainty $\Delta  x$ in position.  For our pendulum $\Delta x \approx \sqrt{\hbar/m \omega} \approx 7.3\times 10^{-18}m$. Thus, even when the pendulum is in a quantum state, for all practical purposes it behaves classically because it has a \emph{macroscopic} mass relative to the accuracy of our imperfect measurement.%
\footnote{For the Heisenberg uncertainty $\Delta  x$  to equal our $\epsilon_{x}  = 10^{-5}$cm,  we would need the mass of the pendulum to be $5.3\times 10^{-25}$kg  (keeping $\omega =1$), and then, even with a single measurement,  we will detect  that the pendulum is not at rest within half a second because its position will change by $\epsilon_{x}$.}
 So long as the measurements are imperfect (i.e.  $\epsilon_{x}  \gg \Delta x$), we can generally ignore the measurement process in the discussion of whether the quantum dynamics of the macroscopic system is well described by its classical description. These simple, order of magnitude considerations will be useful at several junctures. (While we considered a pendulum to obtain explicit numbers, same considerations apply to other macroscopic systems such as a table or the moon.) 

\section{Preliminaries}
\label{s2}

In this section we fix our notation and introduce some mathematical background that will be used in the rest of the paper.  In \ref{s2.1} we recall the notion of the covariant phase space $\Gamma$ for the Klein-Gordon field $\phi$, the symplectic structure $\Omega$ thereon, and the new geometric structure  that one needs to introduce on $\Gamma$ for Fock quantization of $\phi$. In \ref{s2.2}  we introduce the cosmological setting, used in the rest of the paper, and spell out the how the new  structure is chosen in the most commonly used examples -- Minkowski, de Sitter, and quasi-de Sitter space-times, as well as a class of FLRW models that includes radiation and dust filled universes. In \ref{s2.3} we collect the basics of canonical quantization. Notions, structures and expressions introduced in this section are heavily used in the rest of the paper.

\subsection{Geometric Structures on the Covariant Phase space}
\label{s2.1}

Consider a real Klein-Gordon field $\phi$ on a general, globally hyperbolic space-time $(M,\,g_{ab})$ satisfying $\Box \phi =0$.  The \emph{covariant phase space} $\Gamma$ of this field consists of (suitably regular) solutions $\phi$. This space comes equipped with a natural symplectic structure $\Omega$.  Since $\Gamma$ is a vector space, $\Omega$ can be regarded as a skew symmetric (weakly non-degenerate) tensor that associates to any pair  $(\phi_1,\phi_2)$ of vectors in $\Gamma$ a number:
\be \label{sympl}
\Omega (\phi_1,\phi_2) = \int_{\Sigma} \rmd^3V\,  n^{a}\,(\phi_1\nabla_a\phi_2 - \phi_2\nabla_a\phi_1),
\ee
where $\Sigma$ is \emph{any} space-like Cauchy surface and $n^a$ is the unit normal to $\Sigma$. This symplectic structure is induced on $\Gamma$ by a standard procedure starting from the action for the Klein-Gordon field (see, e.g.,  \cite{aa-Lagrange}).

The Fock quantization of the scalar field requires the introduction of a specific new geometric structure on $\Gamma$: a {\it complex structure} $J$ that is compatible with $\Omega$.  More precisely, $J$ is a real-linear mapping, $J:\, \Gamma\to \Gamma$, with the following properties: $J^2=-1$, \, $\Omega(\phi_1,J\phi_2)= \Omega(\phi_2,J\phi_1)$ and $\Omega(\phi,J\phi)\geq 0$ (and it is equal to $0$ if and only if $\phi=0$). When these conditions are satisfied, we can define a {\it positive definite metric} $g$ on $\Gamma$:
\be
g(\phi_1,\phi_2):= \Omega(\phi_1,J\phi_2)\, .
\ee
Thus $(\Gamma, \Omega, J, g)$ is a K\"ahler space.  As we shall see, this extension from a symplectic to a K\"ahler space succinctly captures the passage from the classical to the quantum theory. In particular, $J$ and $g$ have a well defined and precise physical meaning in the quantum theory.

Let us start by considering the complex structure $J$. It provides a splitting of the space of solutions $\Gamma$ into positive and negative frequency solutions: Given any real solution field $\phi$,  we can define its positive and negative frequency parts as:
\be
\phi^{\pm}:= \frac{1}{2}\left(1\mp iJ\right)\phi
\ee
with the property that,
\be
J\, \phi^{\pm}=\frac{1}{2}\left(J\pm i\right)\phi = \pm\frac{i}{2}\left(1 \mp i J\right)\phi=\pm i \phi^{\pm}\, .
\ee
That is, a positive frequency solution $\phi^+$ is an eigenvector of $J$ with eigenvalue equal to $i$ (and, correspondingly, $\phi^-$ with $-i$). Even though $\phi^+$ and $\phi^-$ are both  complex,  $\phi$ is real since $\phi^{-}= \overline{\phi^+}$.   Thus $J$ is the new geometrical structure on the real phase space  $\Gamma$ that captures the notion of positive and negative frequency decomposition. Often this decomposition is accomplished by introducing an appropriately normalized basis ($e_{k}(\eta) e^{i\vk\cdot\vx}$, in the cosmological context discussed below)  of solutions to the Klein-Gordon equation, but the Fock representation of the field algebra depends only on the complex structure $J$ they define; change of the basis leads to the same Fock representation if and only if the change leaves $J$ unchanged.

The second geometrical structure on $\Gamma$  is the positive definite metric $g$ that naturally emerges from the newly introduced complex structure:
\be  \label{metric1} g(\phi_{1},\, \phi_{2}) = \Omega(\phi_{1}, \, J\phi_{2})    \ee
Together with the symplectic structure $\Omega$, the metric $g_{ab}$ enables us to define an inner product on $\Gamma$: 
\be \label{ip1}
\langle \phi_1,\phi_2\rangle:= \frac{1}{2\hbar} g(\phi_1,\phi_2) + \frac{i}{2\hbar} \Omega(\phi_1,\phi_2)\, 
\ee
which is Hermitian if one regards $(\Gamma, J)$ as a complex vector space. Cauchy completion of $(\Gamma, J, \langle.,.\rangle)$ is the 1-particle Hilbert space $\mathcal{H}$ of quantum field theory. (See, e.g., \cite{am, waldbook}).

There is an equivalent description for the 1-particle Hilbert $\mathcal{H}$ space that is more commonly used.  Instead of real solutions $\phi$,  consider their complex-valued positive frequency parts $\phi^+$ which are in 1-1 correspondence with $\phi$. On the space of positive frequency solutions the Hermitian inner product takes the form
\be \label{ip2}
\langle \phi^+_1,\phi^+_2\rangle\, =\, \frac{i}{\hbar} \Omega(\overline{\phi^+_1},\phi^+_2) \,\equiv \, \frac{i}{\hbar} \Omega({\phi}^-_1,\phi^+_2) \, .
\ee
The Hilbert space $\mathcal{H}$ is then the Cauchy completion of the complex vector space spanned by (suitably regular) positive frequency solutions w.r.t. this inner product. The total Hilbert space of the theory is the symmetric Fock space generated by $\mathcal{H}$. { The Fock vacuum is a quasi-free state; its $n$-point functions are determined by the 2-point function.}

Thus, the new element required for quantization is a complex structure $J$ that is compatible with the symplectic structure $\Omega$ on $\Gamma$, such that $(\Omega,\,J,\,g)$ equips $\Gamma$ with a {\it K\"ahler structure}. This geometrical setting for quantization of a scalar field holds on  any globally hyperbolic space-time, not just the FLRW space-times of direct interest to our discussion.%
\footnote{It also admits a simple generalization to spin 1 and spin 2 fields. For fermions,  the role of $\Omega$ and $g$ are reversed: `classical theory' provides us with $g$ and the complex structure then provides us $\Omega$; again it is the K\"ahler structure that defines the Fock quantization (see, e.g., \cite{am-grg}).}  
 
\subsection{FLRW space-times}
\label{s2.2}

Let us now restrict ourselves to the FLRW space-times, 
\be  \label{metric} \mathfrak{g}_{ab}\rmd x^{a} \rmd x^{b} =  a^{2}(\eta)\, \mathring{\mathfrak{g}}_{ab}\rmd x^{a} \rmd x^{b} \equiv
a^{2}(\eta) \big(-\rmd \eta^{2} + \rmd \vx^{2}\big), \ee
so that $\eta$ is the conformal time coordinate, related to proper time $t$ via $a(\eta) \rmd\eta = \rmd t$. To avoid infrared technical complications that are not relevant to our considerations, we will take the spatial topology to be a 3-Torus $\mathbb{T}^{3}$  of (spatial) volume $V_{o}$ with respect to the fiducial flat metric $\mathring{\mathfrak{g}}_{ab}$.  Dynamics of $\phi$ is directly relevant to that of cosmological perturbations in the inflationary scenario, especially for tensor modes.  On this background space-time, the Klein Gordon equation takes the simple form
\be
\phi^{\prime\prime} - \Do^2 \phi + 2 \f{a'}{a}\, \phi' = 0
\label{EOM}
\ee
where prime refers to derivative with respect to conformal time $\eta$, and $\Do$ is the spatial Laplacian defined by $\mathring{\mathfrak{g}}_{ab}$. As is common in the cosmology literature, we will  carry out a Fourier decomposition
\be \label{FT}
\phi(\vx,\eta) = \f{1}{ V_0}\, \sum_{\vk} \, \phi_{\vk} (\eta) \; e^{i\, \vk \cdot \vx}\, .
\ee
%
Because $\phi(\vx,\eta)$ is real, the Fourier transforms are subject to the `reality condition' $\bar{\phi}_{\vk} (\eta) = \phi_{-\vk} (\eta)$. It is customary to introduce a suitably normalized basis $e_{k}(\eta)$ satisfying the equation of motion
\be  \label{EOM2} e_{k}^{\prime\prime} (\eta) + 2 \f{a^{\prime}(\eta)}{a(\eta)}\, e_{k}^{\prime} (\eta) + k^{2} e_{k}(\eta) =0\, ,\ee
and normalization conditions
\be \label{normalization} e_k(\eta)\,\bar{e}^{\prime}_k(\eta)- e^{\prime}_k(\eta)\,\bar{e}_k(\eta)=\frac{i}{a^2(\eta)}, \ee
and expand $\phi_{\vk} (\eta)$ in this basis to obtain 
\be\label{expfield}
\phi(\vx,\eta) = \f{1}{\sqrt{V_{0}}}\,\sum_{\vk} \,(z_{\vk}\, e_k(\eta) + \bar{z}_{-\vk} \, \bar{e}_k(\eta) ) \; e^{i\, \vk \cdot \vx}\, ,
\ee
Here the $z_{\vk}$ are arbitrary complex-valued constants  (subject only to standard fall-off conditions for large $k$ to ensure convergence of the sum in (\ref{expfield})). In particular, in contrast to $\phi_{\vk}$, \emph{there is no relation between $z_{\vk}$ and $z_{-\vk}$}. The $z_{\vk}$ serve as complex (Bargmann) coordinates on the covariant phase space $\Gamma$ \cite{bargmann1,bargmann2}. Note also that the factor of $1/\sqrt{V_{0}}$ has been absorbed in the constants $z_{\vk}$ for later convenience.

The normalization condition (\ref{normalization}) ensures that if one defines a complex structure $J$ on $\Gamma$ via
\be
J\,\phi(\vx,\eta) = \f{1}{\sqrt{V_{0}}}\, \sum_{\vk} \, (i\, z_{\vk}\, e_k(\eta) -i\, \bar{z}_{-\vk} \,\bar{e}_k(\eta) ) \; e^{i\, \vk \cdot \vx}\, .
\ee
then this $J$ is compatible with the symplectic structure $\Omega$. The set  of solutions $\{ e_{k}\, e^{i\vk\cdot\vx} \}$ provides an orthonormal positive frequency basis in the 1-particle Hilbert space $\mathcal{H}$. We can define a new positive frequency basis by replacing $e_{k}$ with  $ \tilde{e}_{k} = \sum_{k^{\prime}} C_{k, k^{\prime}} \,e_{k}$ (with $\sum_{k^{\prime}}\, |C_{k,\, k^{\prime}}|^{2} =1$); the complex structure defined by $\tilde{e}_{k}$ is again $J$. Thus, the invariant content in the choice of a positive frequency basis is captured by the complex structure. It is easy to show for this complex structure, $J$, the Hermitian inner product takes the form,
\be \label{ip3}
\langle \phi_1(\vx,\eta), \phi_2(\vx,\eta)\rangle = \f{1}{\hbar} \sum_{\vk} \, \bar{z}^{(1)}_{\vk} \; z^{(2)}_{\vk}\, .
\ee
Thus, the components $ \{ z_{\vk} \} $ of $\phi(\vx,\eta)$ in the orthonormal basis $e_{k}(\eta)$ provides a convenient coordinate system for $\Gamma$.  Eqs. (\ref{ip2}) and (\ref{ip3}) imply that these coordinates are well-adapted to the symplectic structure: their Poisson brackets have the form
\be \label{pb1} 
\{ z_{\vk},\,\bar{z}_{\vk '}\} = -i\,\delta_{\vk,\vk '}\, \qquad {\rm and} \qquad \{ z_{\vk},\, {z}_{\vk '}\} =0. \ee
\medskip

In the commonly used space-times, the basis $e_{k}(\eta)$ of solutions is chosen as follows.
\begin{enumerate}
\item \emph{Minkowski space-time.} This is of course the simplest homogeneous and isotropic cosmological model, in the sense that it is also stationary. In this case, we have
\be \label{mink-mode}
e_k(\eta)=\frac{e^{-ik\eta}}{\sqrt{2k}}\, 
\ee
where the time dependence is only through the phase factor. These solutions  constitute the standard positive frequency basis. \\

\emph{Remark:} To introduce the mode functions $e_{k}(\eta)$ in FLRW space-times, it is convenient to make a mathematical detour. Recall that in any FLRW model, the equations of motion (\ref{EOM2}) satisfied by the mode functions $e_{k}(\eta)$ imply that  $\chi_{k}(\eta) := a(\eta) e_{k}$ satisfy 
\be \label{chiEOM}
\chi_{k}^{\prime\prime} (\eta) + \big(k^{2} - \f{a^{\prime\prime}}{a}\big) \chi_{k}(\eta) = 0 \ee 
which are generally easier to solve.  In various models of physical interest, one often solves for $\chi_{k}(\eta)$ and then introduces $e_{k}(\eta) = \chi_{k}(\eta)/a(\eta)$ as discussed below.\\

\item \emph{de Sitter space-time.} Here one restricts oneself to the future  Poincar\'e patch, with conformal time $\eta$, such that $a(\eta)=-\frac{1}{H\eta} = e^{Ht}$, where $H$ the constant Hubble parameter and $t$ proper time. In this case the {standard basis functions chosen by appealing to de Sitter isometries} take the form,
\be \label{desittermodes}
e_k(\eta) =  \f{\chi_{k}(\eta)}{a(\eta)} = \frac{1}{a(\eta)}\;\frac{e^{-ik\eta}}{\sqrt{2k}}\, \left( 1 + \frac{iHa(\eta)}{k} \right)\, .
\ee

\item {  \emph{Quasi-de Sitter space-times.} For a single field inflation, consider the dynamical phase in which the Hubble parameter $H := \dot{a}/a$ is not a constant but changes slowly in the sense that the slow-roll parameters $\epsilon:= - \dot{H}/H^{2}$ and $\delta := \ddot{H}/(\dot{H}H)$ are small compared to $1$. (Here the `dot' denotes derivatives with respect to proper time.) It is customary to ignore second and higher order terms in $\epsilon$ and $\delta$. In this approximation, using dynamical equations, the slow roll parameters can be expressed in terms of the inflationary potential $V$:
\be \label{slowroll}
\epsilon\, \approx \,\epsilon_{V} := \f{1}{16\pi G}\, \left(\f{V^{\prime}}{V}\right)^{2} \qquad {\rm and} \qquad \delta\, \approx \,\delta_{V} := \f{1}{8\pi G}\, \f{V^{\prime\prime}}{V} \, ,\ee
where the `prime' now denotes derivative with respect to the (background) inflaton field. The basis adapted to this slow roll phase of dynamics is given by  (see, e.g., \cite{riotto})
\be \label{neardesittermodes}
e_k(\eta)= \f{\chi_{k}(\eta)}{a(\eta)} = - \f{1}{a(\eta)}\,  \frac{\sqrt{\pi}}{2} \; e^{i(\nu+\frac{1}{2})\frac{\pi}{2}}\,\sqrt{-\eta}\; H_\nu^{(1)}(-k\eta)\, ,
\ee
with $\nu=\frac{3}{2}+3\epsilon_{V} -\delta_{V}$ and $H_\nu^{(1)}$ the Hankel function of the first kind.  In the limit $3\epsilon_{V}-\delta_{V} \to 0$ one obtains the de Sitter basis functions. } 

\item \emph{A class of FLRW models interpolating between radiation and dust domination.} Let us begin with a radiation-filled  FLRW universe. In this case, $a(\eta) = a_{0} \eta$, where $a_{0}$ is a constant. Therefore, $a^{\prime\prime} =0$ and hence $\chi_{k}(\eta)$ satisfies the same evolution equation in $\eta$ as the mode functions (\ref{mink-mode})  in Minkowski space-time. Thus, it is now natural to choose as `positive frequency' basis functions
\be \label{radiationmodes}
e_k(\eta) = \f{\chi_{k}(\eta)}{a(\eta)}  = \frac{1}{a_{0} \eta}\; \frac{e^{-ik\eta}}{\sqrt{2k}}\, . \ee
In a dust-filled FLRW universe, we have $a(\eta) = a_{1} \, \eta^{2}$ for some constant $a_{1}$. Interestingly, in this case $\chi_{k}(\eta) := a(\eta) e_{k}(\eta)$ satisfies the same differential equation in $\eta$ as in de Sitter space-time,  $\chi_{k}^{\prime\prime} + (k^{2} -\f{2}{\eta^{2}}) \chi_{k} =0$. Therefore, a natural choice of modes $\chi_{k}(\eta)$ is obtained by multiplying (\ref{desittermodes}) by  the de Sitter scale factor $a(\eta) = -1/H\eta$. Then, for the dust-filled universe  the mode functions $e_{k}(\eta)$ are given by 
\be \label{dustmodes}
e_k(\eta) =  \f{\chi_{k}(\eta)}{a(\eta)}  = \frac{1}{a_{1} \eta^{2}}\; \frac{e^{-ik\eta}}{\sqrt{2k}} \left(1 - \f{i}{k\eta}\right).  \ee
Finally, let us consider FLRW universes that `interpolate' between the radiation and dust filled cases in the sense that the scale factor has the behavior $a(\eta) = a_{\alpha}\, \eta^{(1+\alpha)}$ with $\alpha \in [0,1]$. Then the equation of motion (\ref{chiEOM})  of $\chi_{k} (\eta)$ reduces to $\chi_{k}^{\prime\prime} + (k^{2} -\f{\alpha(1+\alpha)}{\eta^{2}})\, \chi_{k} =0$. Requiring that for large $k$, the positive frequency modes should have the $\eta$-dependence $\sim e^{-ik\eta}$, and noting that $\eta$ is now positive (in contrast to the de Sitter and near-de Sitter space-times), we are led to the mode functions:
\be \label{alphamodes} 
e_{k}(\eta) = \f{\chi_{k}(\eta)}{a(\eta)}  = A_{k,\alpha}\,\f{1}{\eta^{1+\alpha}} \;  \sqrt{k\eta} \,\,H^{(2)}_{\f{1}{2}+\alpha} (k\eta)\, ,  \ee
where $H^{(2)}_{\f{1}{2}+\alpha}(k\eta)$ is the Hankel function of second kind of order $\f{1}{2}+\alpha$, and the normalization constants $A_{k,\alpha}$ are determined using (\ref{normalization}).  For $\alpha=0$ the mode functions (\ref{alphamodes}) reduce to  (\ref{radiationmodes}) with $A_{k,0} = - (i/2a_{0})\, \sqrt{\pi/k}$,  
and for $\alpha=1$ they reduce to (\ref{dustmodes}) with $A_{k,1} = - (1/2a_{1}) \sqrt{\pi/k}$.

This class of models does not feature in most of the discussion of emergence of classicality. It's inclusion will enable us to bring out two interesting features: (i) Inflation is not essential for emergence of semi-classicality in the sense of  Sections \ref{s4} and \ref{s5}; and (ii)  Contrary to a common belief, the three widely used  notions of this emergence are not equivalent. (See also footnote \ref{fn1}.) \\

\emph{Remark:} Since  one generally first solves (\ref{chiEOM}) for $\chi_{k}(\eta)$, one might imagine forgoing the introduction basis functions $e_{k}(\eta)$ altogether and work with the field $\chi_{\vk}(\eta) = \sum_{\vk} (z_{\vk}\, \chi_k(\eta) + \bar{z}_{-\vk} \, \bar{\chi}_k(\eta) )$ in place of $\phi_{\vk}(\eta) = \sum_{\vk} \,(z_{\vk}\, e_k(\eta) + \bar{z}_{-\vk} \, \bar{e}_k(\eta) )$. However, it is $\phi_{\vk}$ that is directly need in physical applications. In particular, the primordial TT-power spectrum is the 2-point function of $\hat\phi_{\vk}$, not of $\hat\chi_{\vk}$. Similarly it is the observable $\hat{\phi}_{k}$ that is squeezed during inflation; in fact the uncertainty in $\hat{\chi}_{\vk}$ increases exponentially in the number of e-folds during inflation.  Therefore,  physical considerations involving emergence of classical behavior refer to the field $\hat{\phi}_{\vk}$.


\end{enumerate}

\subsection{Canonical Commutation Relations}
\label{s2.3}
Let us now introduce the canonical variables that will play an important role  in the following sections. In the Fock representation defined by the complex structure $J$ associated with a basis $e_{k}$,  the final result amounts to replacing the coefficients $z_{\vk}$ in Eq(\ref{expfield}) with $\ak$. Thus, we have:
\be \label{phiop}
\hat{\phi}(\vx,\eta) = \f{1}{\sqrt{{V_{0}}}} \,\sum_{\vk} \,(e_k(\eta)\ak + \bar{e}_k(\eta)\admk ) \; e^{i\, \vk \cdot \vx}\, ,
\ee
where the creation and annihilation operators are subject to the commutation relations
\be
[ \ak, \adag_{\vk^{\prime}} ] = \hbar\, \delta_{\vk,\vk^\prime}\, \qquad {\rm and} \qquad  [ \ak, \hat{A}_{\vk^{\prime}} ]  =0\,  \ee
that mirror the Poisson brackets (\ref{pb1})  between $z_{\vk}$ and $\bar{z}_{\vk}$. The Fock vacuum $|0\rangle$ is defined by $\hat{A} |0\rangle =0$.  For \emph{any} choice of basis $e_{k}$ satisfying the equations of motion (\ref{EOM}) and the normalization condition (\ref{normalization}) the vacuum is invariant under spatial translations and rotations of the FLRW space-time. These vacua are all quasi-free states and hence determined by their 2-point functions. The invariance of the vacuum under these isometries is therefore equivalent to that of the 2-point function 
\be \langle \hat\phi(\vx_{1}, \eta_{1}) \, \hat\phi (\vx_{2}, \eta_{2}) \rangle =  \f{\hbar}{V_{0}}\, \sum_{\vk}\, e_{k}(\eta_{1})\, \bar{e}_{k}(\eta_{2}) \, \, e^{i{(\vec{x}_{1} - \vec{x}_{2})\cdot\vk}}  \ee
which is manifest, by inspection, for any choice of basis functions.

Fix an instant $\eta_{o}$ of time. Then the canonically conjugate pair of operators at $\eta_{o}$ is given by
\be \hat\varphi(\vx) = \hat\phi(\vx, \eta_{o})  \qquad {\rm and}\qquad  \hat\pi(\vx) =  a^{2}(\eta_{o}) \, \hat{\phi}^{\prime}(\vx, \eta_{o}),  \ee
where, as before, prime refers to the derivative with respect to $\eta$.
Thus, we have expansions:
\be \label{op-phi}  \hat{\varphi}(\vx, \eta) = \f{1}{\sqrt{V_{0}}} \,\sum_{\vk} \,(e_k(\eta)\ak + \bar{e}_k(\eta)\admk ) \; e^{i\, \vk \cdot \vx}\, =:  \f{1}{\sqrt{V_{0}}} \,\sum_{\vk} \,  \hvp_{\vk}\, e^{i\, \vk \cdot \vx}
\ee
and
\be\label{op-pi}
\hat{\pi}(\vx,\eta) = \f{a^2(\eta)}{\sqrt{V_{0}}}\, \sum_{\vk} \,(e'_k(\eta)\ak + \bar{e}'_k(\eta)\admk ) \; e^{i\, \vk \cdot \vx}\, =: \f{1}{\sqrt{V_{0}}} \,\sum_{\vk} \,  \hat\pi_{\vk}\, e^{i\, \vk \cdot \vx}\,  \ee
so that  $\hvp^{\dag}_{\vk} = \hvp_{-\vk}$ and  $\hat\pi^{\dag}_{\vk} = \hat\pi_{-\vk}$\,. The commutation relations between $\ak$ and $\adk$ then imply the canonical commutation relations:
\be [\hat\varphi(\vx),\,  \hat{\pi}(\vx^{\prime}) ] = i\hbar \delta(\vx,\, \vx^{\prime}) \quad {\rm and} \quad  [\hvp_{\vk},\, \hat\pi_{\vk^{\prime}}] = i\hbar \delta_{\vk, \, -\vk^{\prime}}\, .
\ee

\section{Quantum non-commutativity and inflation}
\label{s3}

In this section we address the following questions: Is there a precise sense in which non-commutativity `fades' during inflation? And, if so, is such fading a good criterion for emergence of classical behavior?

\subsection{ Strategy}
\label{s3.1}
Let us begin by spelling out the general context. As in section \ref{s2}, consider a general FLRW background and use suitable mode functions $e_{k}(\eta)$ to define the complex structure,  and consider the resulting Fock representation of the operator algebra. Then,  the Fourier transforms of the canonically conjugate pair of operators are given by
\be \label{phipi} \hvpk (\eta)=  e_{k}(\eta) \ak+ \bar{e}_{k}(\eta) \admk \quad{\rm and} \quad \hpik(\eta)  = { a^{2}(\eta)} \big(e_{k}^{\prime}(\eta) \ak+ \bar{e}^{\prime}_{k}(\eta) \admk\big)  \ee
at any conformal time $\eta$, and that they satisfy the canonical commutation relations 
\be \label{CCR2} [\hvpk(\eta),\, \hpikp(\eta) ] = i\hbar\, \delta_{\vec{k}, -\vkp}\,\,.   \ee
This non-commutativity is of course a key hallmark of quantum mechanics. Therefore, if it were to become negligible in an appropriate sense during dynamics, one could say that the system exhibits classical behavior in that phase of evolution.

Now, as we saw in Section \ref{s2}, in de Sitter space-time the `positive frequency' basis functions are given by:
\be \label{BD} e_{k}(\eta)  = \Big(\f{1}{a(\eta)}  + i \f{H}{k}\Big) \, \f{e^{-ik\eta}}{\sqrt{2k}},  \ee
It is sometimes argued that, since  $a(\eta)$ becomes very large at late times, the `decaying mode' would become negligible and then the canonically conjugate operators would `approximately \emph{commute} at late times'.%
\footnote{Using the same logic, it is argued that the field operators $\hphik(\eta_{1})$ and $\hat\phi_{\vk^{\prime}}(\eta_{2})$ also ``approximately commute at different times'' provided $\eta_{1}$ and $\eta_{2}$ are taken to be sufficiently late (see, e.g., \cite{kiefer2}).  This issue is discussed in Appendix \ref{a1}. We will see that {while this expectation is not borne out as stated,  it does hold in the sense spelled out in this sub-section. The strategy introduced in this sub-section also brings out some interesting features in the \emph{way} in which the non-commutativity of $\hphik(\eta_{1})$ and $\hat\phi_{\vk^{\prime}}(\eta_{2})$ `fades'  that, to our knowledge, have not been noticed before.} }
As it stands, this reasoning is incorrect because,  as Eq.(\ref{CCR2}) shows,  the commutator between $\hvpk(\eta)$ and  $\hpikp(\eta)$ is time independent, whence it is the same at late times as it was at early times. Nonetheless, one can ask whether properties of the basis functions $e_{k}(\eta)$ can lead to `fading of non-commutativity' in some well-defined sense. 

In this section we will answer this question affirmatively in the inflationary context. To do so let us first note that, since operators involved are all unbounded, it is not meaningful to say that one part of the operator becomes negligible; one can always find states on which it is far from being so.  Secondly, the commutator is dimensionfull, whence it can be compared to --and then regarded as negligible-- only with respect to a quantity that has the same physical dimensions.  A natural strategy is to compare the \emph{expectation value} of the commutator between $\hvpk(\eta)$ and  $\hpikp(\eta)$  with that of the \emph{anti-commutator} between the same operators. One would then compare numbers --rather than operators-- both of which have the same physical dimension.  The question would be whether the expectation value of the commutator becomes small compared to that of the anti-commutator under time evolution. 

This strategy can be motivated by two considerations. The first comes from results presented in section \ref{s5} which imply that 
the expectation value of the anti-commutator equals a classical quantity involving the two observables. On the other hand the expectation value of the commutator is a quintessentially quantum quantity. Therefore, the ratio of the two expectation values can be taken to be a measure of the `importance of the quantum aspects of the system relative to its classical aspects'.  A second and independent motivation comes from the structure of the algebra of observables in classical and quantum mechanics \cite{aa-cmp}. For quantum mechanical systems whose configuration space $\mathcal{C}$ is a manifold, it is natural to associate configuration observables with functions on $\mathcal{C}$ and momentum observables with vector fields on $\mathcal{C}$. These two sets of observables are subject to certain algebraic relations. It turns out that the classical and quantum algebras share the \emph{same anti-commutation relations}. But the commutation relations are of course different: Classical observables commute, while the quantum observables do not. This structure also suggests that the ratio we consider is a measure of the quintessentially quantum behavior.

We will find that, with this specific formulation not only does the question become well-defined but the intuitive idea of `fading of non-commutativity' is realized as inflation unfolds.  However, we will also find that this criterion of classical behavior has important limitations, illustrated by the explicit example of quantum fields propagating on a radiation-filled FLRW space-time. The criteria discussed in Sections \ref{s4} and especially \ref{s5} are better suited to capture the idea of emergence of classical behavior in more general circumstances.

\subsection{The canonically conjugate operators at any given time $\eta$}
\label{s3.2}

Let us begin with a general FLRW space-time and work with the Fock representation defined by a given set $\{e_{k}(\eta)\}$ of (`positive frequency') basis functions. As is common in cosmology, we will work in the Heisenberg picture and use as our state the  vacuum that is annihilated by the operators $\ak$ of Eq. (\ref{phipi}).  Then a straightforward calculation shows that, at any time $\eta$, the vacuum expectation values of the commutators and anti-commutators of the pair $(\hvpk,\, \hpik )$ of operators are given by
\ba
&\langle\,\,[\hvpk (\eta),\, \hpi_{\vk'}(\eta)]\,\, \rangle=\, i \,\hbar  \delta_{\vec{k} , -\vkp}\notag \\
&\langle\,\, [\hvpk (\eta), \, \hpi_{\vk'}(\eta)]_{+}\,\,\rangle = \,  - \,2\hbar\, a^{2}(\eta)\,{\rm Re} (e_{k}(\eta) \, \bar{e}_{k^{\prime}}^{\prime})\,\, \delta_{\vec{k} , -\vkp}\, .
\ea
Therefore the absolute value $|R_{\vp,\pi}|$ of the ratio of the expectation value of the commutator to that of the anti-commutator is of interest only if $\vec{k} = - \vec{k}^{\prime}$ and is then given by 
\be\big{ |}R_{\vp,\pi} (\eta) \big{|} \,:=\, \Big{|}\f{\langle\,\,[\hvpk (\eta),\, \hpi_{-\vk}(\eta)]\,\, \rangle}{\langle\,\, [\hvpk (\eta), \, \hpi_{-\vk}(\eta)]_{+}\,\,\rangle} \Big{|}\, =\, \f{1}{{\big{|}2\, {a^{2}(\eta)} \rm Re}\, (e_{k}(\eta) \, \bar{e}_{k}^{\prime}(\eta))\big{|}} \, \ee
Let us now specialize to (the future Poincar\'e patch of) de Sitter space-time and use for $e_{k}(\eta)$ the basis functions (\ref{BD}). Then  $ |R_{\vp,\pi}|$ simplifies:
\be \big{|}R_{\vp,\pi} (\eta) \big{|} \,=\, \f{k}{H} \, \f{1}{a(\eta)} \, =\, \f{k_{\rm phy} (\eta)}{H} \, \ee
where $k_{\rm phy}$ is the physical wave-number. Let us denote by $\eta_{k}$ the time when the mode exits the Hubble horizon, i.e.,  when $k_{\rm phy}= H$, and $\eta$ the time $N$  e-folds later.  Then we have:
\be \big{|} R_{\vp,\pi} (\eta_{k}) \big{|}  =  1\quad {\rm and}\quad  \big{|}R_{\vp,\pi} (\eta) \big{|} = \, e^{-N} \, .\ee
Thus,  neither the commutator nor the ratio is negligible at the horizon crossing time $\eta_{k}$. But while the commutator is independent of $\eta$, the ratio decreases exponentially with the number $N$ of e-folds as inflation proceeds. The smallest wavelength mode observed by the Planck satellite exits the Hubble horizon $\sim 8$ e-folds after the longest wavelength mode, and  there are $\sim 55$ e-folds in the relevant phase of the slow roll (see footnote \ref{fn1}). Therefore,  $N=8$  and $N = 55$ are useful number to keep in mind.  Already 8 e-folds after $\eta_{k}$, the ratio $ | R_{\vp,\pi}|$ is reduced by a factor of $\sim 3.3 \times 10^{-4}$, and for 55 e-folds by $\sim 1.3 \times 10^{-24}$;  this can be taken as a precise sense in which the non-commutativity between the field operator and its conjugate momentum diminishes after horizon crossing.

To summarize, the commutator $[\,\hvpk (\eta),\, \hpi_{\vk'}(\eta)\,]$ --and its expectation value in any state-- is time independent; it does not decay. However, the ratio of the expectation values of the commutator and the anti-commutator in the Bunch-Davies vacuum decays exponentially with the number of e-folds to the future of horizon-crossing, providing us with a precise sense in which the significance of non-commutativity `fades'. This occurs because the expectation value of the anti-commutator is proportional to the scale factor in de Sitter space-time. \\
\goodbreak
\emph{Remarks:}

{ 1. The situation in quasi-de Sitter space-times is similar because the mode functions approximate those in de Sitter space-time quite well. More precisely,  in comparison with the de Sitter space-time, there are only two notable modifications: (i)  $H := \dot{a}/a$
is no longer a constant, but varies slowly, whence the scale factor now has a form $a(\eta) = (-\mathring{H}\eta)^{-(1+\epsilon_{V})}$ for some constant $\mathring{H}$; and, (ii)  at late times  the mode functions $e_{k}(\eta)$ have an additional time dependence  of  $(-\eta)^{\delta_{V} -2\epsilon_{V}}$ (see, e.g., \cite{hks}).   Since $\delta_{V}$ and $\epsilon_{V}$ are small during slow roll, the denominator in the ratio $\big{|} R_{\vp,\pi} (\eta_{k}) \big{|}$ continues to grow as in de Sitter space-time and so non-commutativity again fades. (Incidentally, note that $\delta_{V} = 2\epsilon_{V}$ for the quadratic potential, whence the extra time dependence disappears there.)}

2. In the above analysis we focused our attention on just two modes $\vk$ and $-\vk$.
However,  the reasoning can be extended to the full Klein-Gordon field $\hvp (\vx, \eta)$ and its conjugate momentum $\hat\pi(\vx, \eta)$ if one phrases the question of  fading of non-commutativity appropriately. One would now tailor the question to a \emph{subset} of configuration and momentum observables: $(\hat\varphi(f)) (\eta)$  and $(\hat\pi(g)) (\eta)$, obtained by smearing $\hvp (\vx, \eta)$ and  $\hat\pi(\vx, \eta)$ with suitable test functions $f(\vx)$ and $g(\vx)$. If these test functions are chosen so that their Fourier transforms have support only on a finite but arbitrarily large band of modes, it again follows that the ratio of the expectation values of commutators and anti-commutators of the resulting set of observables decays exponentially with the number of e-folds after the mode with the largest value of $k$ in the band exits the Hubble horizon. However, for the full algebra of observables generated by all permissible test fields, we do not have a simple statement of fading.

3. Although the importance of non-commutativity does fade during inflation in a well-defined sense,   examination in more general contexts beyond inflation shows that this fading is not a robust criterion for the emergence of classical behavior, because there are situations in which the expectation value of the anti-commutator may become very small --making the ratio large. This occurs in the radiation or dust filled universes as well as the 1-parameter family of cases  labelled by $\alpha \in [0,1]$ that interpolate between the two in the sense of Section \ref{s2.2}. Properties of the Hankel functions $H^{(2)}_{\f{1}{2}+\alpha}$ in the basis functions $e_{k}(\eta)$ of (\ref{alphamodes}) immediately imply that $a^{2}(\eta)  {\rm Re}\, (e_{k}(\eta)\, \bar{e}_{k}^{\prime}(\eta))$ decays as $\eta$ increases. Therefore the ratio $R$ grows as the universe expands,  whence the non-commutativity does \emph{not} fade. However, in Section \ref{s4} we will find that not only is there squeezing in $\hvpk$ in these space-times, but in a precise sense it is even more pronounced than the one resulting from inflation! Similarly, in this case, classical behavior does emerge in the sense of Section \ref{s5}.  Thus, in cosmological contexts beyond inflation, the three notions of  emergence of classical behavior are distinct.

Indeed, even in a dynamical phase where the system behaves classically in an `obvious' physical sense, non-commutativity need not fade. Let us consider the simple example of the grandfather clock from Section \ref{s1}. Suppose it is well isolated so the effect of environment is completely negligible and we do not bring in decoherence. Still, as we argued, this macroscopic system exhibits classical behavior if the pendulum is in its ground state, as can be checked by carrying out (imperfect) measurements of its position repeatedly.  Does the quantum mechanical non-commutativity fade in this case? Again, the canonical commutation relations are time independent; they do not fade. What about the ratio of expectation values of the commutator and the anti-commutator? Can it be used as a pointer that anticipates classical behavior?  Unfortunately, the ratio is infinite because the expectation value of the commutator is just $i\hbar$  while that of the anti-commutator is zero.  { This system does not exhibit classical behavior in the sense of squeezing of Section \ref{s4} either, although it does satisfy the classicality criterion of Section \ref{s5}.}

\section{Phase Space, Quantization and Geometry of Squeezing}
\label{s4}

The phenomenon of quantum squeezing is often used to argue that classical behavior naturally emerges during inflation because the uncertainty in the field configuration $\hvpk$  is highly squeezed  and remains squeezed at late times. In Section \ref{s4.1}  we set the stage by briefly recalling this well-known phenomenon in the context of de Sitter space-time (see, e.g., \cite{lpg,albretcht,dpaas,lps,kiefer1,kiefer2,jmvv1}). { In Section \ref{s4.2}  we trace back the origin of quantum squeezing to geometrical  structures on the classical phase space, introduced in Section \ref{s2.1}. As a result, we will find that the phenomenon is rather general and inflation is not essential for its occurrence.}

\subsection{Squeezing during inflation}
\label{s4.1}

Consider the operators $\hat{\varphi}_{\vk}(\eta)$ and $\hat{\pi}_{\vk}(\eta)$ of Eqs (\ref{op-phi})  and  (\ref{op-pi}). Their expectation value in the vacuum state 
selected by the basis functions $e_{k}(\eta)$ vanishes, $\langle \hat{\varphi}_{\vk}(\eta) \rangle = \langle \hat{\pi}_{\vk}(\eta) \rangle = 0$, so the uncertainties are given by the 2-point functions,
\be
|\Delta\hat{\varphi}_{\vk}(\eta)|^2= \langle \hat{\varphi}^{\dagger}_{\vk}(\eta)\,\hat{\varphi}_{\vk}(\eta) \rangle
\label{unsphi}
\ee
and
\be
|\Delta\hat{\pi}_{\vk}(\eta)|^2= \langle \hat{\pi}^{\dagger}_{\vk}(\eta)\,\hat{\pi}_{\vk}(\eta) \rangle\, .
\label{unspi}
\ee
These uncertainties can be computed using the expansion (\ref{op-phi}) and (\ref{op-pi})  of $\hat{\varphi}_{\vk}(\eta)$ and $\hat{\pi}_{\vk}(\eta)$ in terms of creation and annihilation operators:
\be
\langle \hat{\varphi}^{\dagger}_{\vk}(\eta)\,\hat{\varphi}_{\vk}(\eta) \rangle = \hbar\, e_k(\eta)
\,\bar{e}_k(\eta)\, ,
\label{2-point1}
\ee
and,
\be
\langle \hat{\pi}^{\dagger}_{\vk}(\eta)\,\hat{\pi}_{\vk}(\eta) \rangle = \hbar\,a^4(\eta)\, e^{\prime}_k(\eta)
\,\bar{e}^{\prime}_k(\eta)\, .
\label{2-point2}
\ee

{ Let us evaluate these expressions in de Sitter space-times. Recall that the basis functions are given by} $e_k(\eta)=\frac{e^{-ik\eta}}{\sqrt{2k}}\; \frac{1}{a(\eta)}\left( 1 + \frac{iHa(\eta)}{k}\right)$. Therefore in the Bunch-Davies vacuum the uncertainties are given by:
\be \label{phi-uncertainty}
|\Delta \hat{\varphi}_{\vk}(\eta)|^2 = \frac{\hbar}{2k}\left( H^2\eta^2 + \frac{H^2}{k^2} \right) = \frac{\hbar}{2k}\left( \frac{1}{a^2(\eta)}+ \frac{H^2}{k^2} \right)\, .
\ee
and 
\be  \label{pi-uncertainty}
 \langle \hat{\pi}^{\dagger}_{\vk}(\eta)\,\hat{\pi}_{\vk}(\eta) \rangle = \hbar \f{k}{2} \, \f{1}{H^{2}\eta^{2}}  = \hbar \f{k}{2}\, a(\eta)^{2}
\ee
Let us track how these uncertainties change in time.   We are interested in modes that are deep inside  the Hubble horizon at early times, i.e., satisfy $k_{\rm phy} = {k}/{a(\eta)} \gg H$ then. Let us fix the convention that at an early  proper (or cosmic) time $t=0$, the scale factor is given by $a=1$.  Then modes of interest satisfy $k \gg H$ and we have
\be
|\Delta \hat{\varphi}_{\vk}(\eta)|^2 = \frac{\hbar}{2k}\left( 1+ \frac{H^2}{k^2} \right) \,\approx\,   \frac{\hbar}{2k}; \qquad |\Delta\hat{\pi}_{\vk}(\eta)|^2= \hbar \frac{k}{2}\, .
\ee
These are precisely the uncertainties associated with the vacuum state in Minkowski space-time.
In particular, their product  is (nearly) saturated and uncertainties are ``as equally distributed" as dimensional consideration allow.

{Let us now examine what happens to the modes under consideration at late times} --several e-folds after the mode has crossed the Hubble horizon so $k_{\rm phy} (\eta) \equiv k/a(\eta) \ll H$.
Then (\ref{phi-uncertainty}) simplifies and we have:
\be
|\Delta \hat{\varphi}_{\vk}(\eta)|^2 \approx \frac{\hbar}{2k}\, \frac{H^2}{k^2} \, ,
\ee
which approaches a constant. On the other hand, the uncertainty (\ref{pi-uncertainty})  in the canonically conjugate momentum takes the form
\be
|\Delta\hat{\pi}_{\vk}(\eta)|^2= \hbar \,\frac{k}{2}\, a^2(\eta) \, ,
\ee
that grows unboundedly. Thus, now the product of uncertainties is far from being saturated and grows exponentially with the number of e-folds. However, at late times the uncertainty in $\hvpk$ is very small compared to that in Minkowski space-time because $k\gg H$. This is the manifestation of quantum squeezing during inflation.

\subsection{Geometry of Squeezing}
\label{s4.2}

The phenomenon of squeezing in the early universe is quintessentially quantum since it refers to the evolution of uncertainties in canonically conjugate observables in a given vacuum. Now, we saw in Sec.~\ref{s2} that the passage from classical to quantum theory of linear fields can be systematically streamlined: The symplectic geometry on the classical phase space $\Gamma$ has to be extended to a K\"ahler geometry by introducing a complex structure $J$ --or equivalently a positive definite Riemannian metric $g$-- on $\Gamma$, that is compatible with the symplectic structure $\Omega$ thereon. Therefore, one might expect that the phenomenon of squeezing can be traced back to the classical phase space $\Gamma$ once it is equipped with an appropriate metric $g$. We will now show  that this expectation is indeed correct. The analysis will bring out an interesting and rather unforeseen interplay between the symplectic and Riemannian geometries on $\Gamma$. It will also serve to bring out the fact that inflation is not essential for this phenomenon to occur.

Recall from sections \ref{s2} and \ref{s3} that the phase space $\Gamma$ admits two sets of convenient (complex-valued) canonically conjugate coordinates. The first --Bargmann variables--  are given by $z_{\vk}$ and $\bar{z}_{\vk}$, where $z_{\vk}$ are freely specifiable (apart from appropriate fall-off conditions for large $k=|\vk|$). The second set is provided by the pair $(\varphi_{\vk}(\eta),\bar{\pi}_{\vk}(\eta))$ for any fixed time $\eta$, and subject to the reality conditions $\bar{\varphi}_{\vk} (\eta) = \varphi_{- \vec k}(\eta)$ and $\bar{\pi}_{\vk} (\eta)= \pi_{- \vec k}(\eta)$. Recall that these pairs have the Poisson bracket relations:
\be \label{pb2}
\{ z_{\vk}, \bar{z}_{\vk '} \} = -i\,\delta_{\vk,\vk '}\qquad {\rm and} \qquad \{ z_{\vk},{z}_{\vk '} \} = 0 \, ,
\ee
for the Bargmann variables and 
\be \label{pb3}  \{ \varphi_{\vec{k}}(\eta), \pi_{\vec{k'}}(\eta) \} = \delta_{\vec{k}, -\vec{k'}}\,, \quad{\rm and} \quad \{ \varphi_{\vec{k}}(\eta), \varphi_{\vec{k'}}(\eta) \} =0\, \quad 
\{ \pi_{\vec{k}}(\eta), \pi_{\vec{k'}}(\eta) \} = 0\, ,
\ee
for the canonically conjugate pairs at any fixed time $\eta$.

Recall that every function $f$ on $\gamma$ defines a Hamiltonian vector field  (HVF)%
\footnote{Note that in the mathematical terminology a  `Hamiltonian vector field' is associated with every phase space function $f$; not just with the physical Hamiltonian of the system.}
$X_{f}$ via 
\be
{\rmd} f= \Omega(\cdot, X_f)\, , \qquad \hbox{\rm or, equivalently} \qquad
X^\alpha_f=\Omega^{\alpha\beta}\partial_{\beta}f
\ee
where $\alpha, \beta \ldots $ are abstract indices (\'a la Penrose) that refer to the tangent and co-tangent space of $\Gamma$ \cite{pr,ahm}. Of special interest are the HVFs associated with phase space coordinates  --such as the pairs  $(z_{\vk}, \bar{z}_{\vk})$ and $(\vpk, \pi_{\vk})$\,--
because they provide a convenient basis in the tangent space of $\Gamma$. In particular, the vector fields
 \begin{equation}
X_{z_{\vk}} = i \frac{\delta}{\delta \bar{z}_{\vk}} \,\,\, {\rm or,} \,\,\,X_{z_{\vk}}^{\alpha} =  i \, \bar{Z}_{\vk}^{\alpha};\quad {\rm and} \quad X_{\bar{z}_{\vk}} = -i \frac{\delta}{\delta z_{\vk}} \,\,\, {\rm or,} \,\,\,X_{\bar{z}_{\vk}}^{\alpha} =  -i \, {Z}_{\vk}^{\alpha};
\end{equation}
span the (complexified)  tangent space of $\Gamma$ and their symplectic inner products 
\be
\Omega(X_{\bar{z}_{\vk}},\, X_{z_{\vk^{\prime}}}) =i \delta_{\vec{k}, \vec{k'}}\qquad
{\mathrm{and}} \qquad
\Omega(X_{z_{\vk}} ,\, X_{z_{\vk^{\prime}}}) = 0
\ee
simply reflect the Poisson bracket relations (\ref{pb2})  because $\{ z_{\vk}, \bar{z}_{\vk^{\prime}} \}\equiv \Omega( X_{z_{\vk}},\, X_{\bar{z}_{\vk^{\prime}}})$.

Recall from Section \ref{s2.1} that the passage to the quantum theory requires the introduction of  a new structure on $\Gamma$, namely a complex structure ${J^{\alpha}}_{\beta}$ that is compatible with the symplectic structure $\Omega$, and that the Bargmann coordinates $z_{\vk}$ provide such a ${J^{\alpha}}_{\beta}$. Its action on the Bargmann HVF is given by
\be
{J^{\alpha}}_{\beta}\, Z^{\beta}_{\vk}= i Z^{\alpha}_{\vk}\qquad ; \qquad {J^{\alpha}}_{\beta}\, \bar{Z}^{\beta}_{\vk}= - i \bar{Z}^{\alpha}_{\vk}\, .
\ee
The second structure is a metric { $g_{\alpha\beta} := \Omega_{\gamma\alpha} \,J^{\gamma}_{\beta}$} on $\Gamma$ that allows us to define (Riemannian) inner products between tangent vectors. For the Bargmann vector fields we obtain
\be
g(X_{\bar{z}_{\vk}},\, X_{z_{\vk^{\prime}}}) = \delta_{\vec{k}, \vec{k'}}\qquad
{\mathrm{and}} \qquad g(X_{z_{\vk}},\, X_{z_{\vk '}}) = 0\, .
\ee
Thus, the HVFs associated with Bargmann coordinates provide us with a {(null) complex-valued orthogonal} basis on $\Gamma$.\\

\emph{Remark:} We can also introduce \emph{real} canonically conjugate coordinates {  $(q_{\vk},p_{\vk})$ via  $\sqrt{2} \,z_{\vk} = q_{\vk}\, +\, i\,p_{\vk}$,\, so that $\{q_{\vk},\, p_{\vk^{\prime}}\} = \delta_{\vk, \vk^{\prime}}$. }The corresponding real Hamiltonian vector fields $X_{q_{\vk}}$ and $X_{p_{\vk}}$ then satisfy:
\be \label{simpip}
\Omega(X_{q_{\vk}}, X_{p_{\vk^{\prime}}})=  \delta_{\vec{k}, \vec{k^{\prime}}} \quad {\rm and}  \quad \Omega(X_{q_{\vk}}, X_{q_{\vk^{\prime}}})= 0;  \quad \Omega(X_{p_{\vk}}, X_{p_{\vk^{\prime}}})= 0\, .
\ee
and,
\be  \label{ip}g (X_{q_{\vk}}, X_{q_{\vk ^{\prime}}}) = \delta_{\vec{k}, \vec{k^{\prime}}}; \quad g (X_{p_{\vk}}, X_{p_{\vk ^{\prime}}}) = \delta_{\vec{k}, \vec{k^{\prime}}}; \quad g (X_{q_{\vk}}, X_{p_{\vk ^{\prime}}}) = 0 \ee
Thus, the HVFs defined by  $(q_{\vk},p_{\vk})$ provide us with a real orthonormal basis on the phase space that can be used in place of the one provided by $(z_{\vk}, \bar{z}_{\vk}$).\\

Let us now consider the one parameter family $(\varphi_{\vk}(\eta),\pi_{-\vk}(\eta) \equiv \bar\pi_{\vk}(\eta))$ of phase space functions parametrized by conformal time $\eta$ that have the direct interpretation as the field and its canonically conjugate momentum at time $\eta$.%
\footnote{Recall that if $\phi(\vx, \eta)$ represents scalar/tensor cosmological perturbations, the power spectrum is given by the 2-point function, and the bi-spectrum is related to the 3-point function, both constructed from $\hvpk$.} 
Let us now find expressions for the Hamiltonian vector fields of these coordinate functions on $\Gamma$. Since
\be
\varphi_{\vk}(\eta) = e_{k}(\eta) z_{\vk} + \bar{e}_{k}(\eta) \bar{z}_{-\vk}\, ,
\ee
and
\be
\bar\pi_{\vk}(\eta) = a^2(\eta)\, ( \bar{e}'_{k}(\eta) \bar{z}_{k} + e'_{k}(\eta) z_{-\vk} )\, ,
\ee
it follows that,
\be
X_{\varphi_{\vk}(\eta)} = e_{k}(\eta) X_{z_{\vk}} + \bar{e}_{k}(\eta) X_{\bar{z}_{-\vk}}\, ,
\ee
and
\be
X_{\bar\pi_{\vk}(\eta)} = a^2(\eta)\, ( \bar{e}'_{k}(\eta) X_{\bar{z}_{k}} + e'_{k}(\eta) X_{z_{-\vk}} )\, .
\ee
Again, the symplectic inner products between these HVFs, 
\be \label{sympip2}
\Omega(X_{\varphi_{\vk}(\eta)}, X_{\bar\pi_{\vk^{\prime}}(\eta)} ) = -\delta_{\vec{k}, \vec{k'}}; \quad {\rm and}\quad 
\Omega(X_{\varphi_{\vk}(\eta)}, X_{\varphi_{\vk'}(\eta)} ) = 0; \quad
\Omega(X_{\bar\pi_{\vk}(\eta)}, X_{\bar\pi_{\vk'}(\eta)} ) = 0\, ,
\ee
simply reflect the Poisson Bracket relations (\ref{pb3})  between $\varphi_{\vk}(\eta)$ and $\pi_{\vk}(\eta)$.\medskip

{ With these preliminaries out of the way,}  we can now use these vector fields to explore the physical information contained in the Riemannian metric $g$ --the new geometric structure on $\Gamma$  that is necessary for passage to the quantum theory. From geometric perspective,  the role of a Riemannian metric is to define inner products between vectors, and in particular their norms. We have a set of natural vector fields, namely the HVF associated to field and momenta coordinates evaluated at any instant of time $\eta$. Let us compute their norms:
\be \label{Eq-Imp1}
g(\bar{X}_{\varphi_{\vk}(\eta)}, X_{\varphi_{\vk}(\eta)} ) = 2\, e_{k}(\eta) \,\bar{e}_{k}(\eta)\, ,
\ee
and
\be \label{Eq-Imp2}
g(\bar{X}_{\bar\pi_{\vk}(\eta)}, X_{\bar\pi_{\vk}(\eta)} ) = 2\,a^4(\eta)\, e'_{k}(\eta) \,\bar{e}'_{k}(\eta)\, ,
\ee
and examine how they `evolve' as $\eta$ changes. These expressions have two interesting features.
\begin{enumerate}
\item While $\varphi_{\vk}(\eta)$, $\bar\pi_{\vk}(\eta)$ depend explicitly on the conformal time $\eta$, the symplectic inner product (\ref{sympip2})  between their Hamiltonian vector fields is $\eta$-independent, reflecting the fact that the dynamical flow that evolves this canonically conjugate pair preserves the symplectic structure. However, it does not preserve the complex structure, nor the metric. Consequently, the norms of these Hamiltonian vector fields are explicitly time dependent on general cosmological backgrounds. 
\item Furthermore, the time dependence is such that  the norm of $X_{\varphi_{\vk}(\eta)}$ is precisely the 2-point function calculated in (\ref{2-point1}), while the norm of $X_{\pi_{\vk}(\eta)}$ is given by the 2-point function (\ref{2-point2}), both up to a factor of $\hbar/2$ (that arises for dimensional reasons and conventions):
\be
|\Delta \varphi_{\vec{k}}| = \frac{\hbar}{2}\, g(\bar{X}_{\varphi_{\vk}(\eta)}, X_{\varphi_{\vk}(\eta)} ) \qquad ; \qquad
|\Delta \pi_{\vec{k}}| = \frac{\hbar}{2} \, g(\bar{X}_{\pi_{\vk}(\eta)}, X_{\pi_{\vk}(\eta)} ) \, .
\ee
As a consequence, \emph{the origin of the quantum phenomenon of squeezing can be directly traced back to the fact that the Riemannian metric $g$ on $\Gamma$ fails to be preserved by the dynamical flow.}%
\footnote{\label{fn7}There is  an alternative description of this geometric origin of squeezing. Consider the canonical phase space $\Gamma_{\!{}{\rm can}}$, consisting of fields $(\varphi(\vx), \pi(\vx))$ on $\mathbb{R}^3$. Using Fourier transforms, it can be coordinatized by the pairs $(\varphi_{\vk}, \bar{\pi}_{\vk})$ satisfying the reality conditions $\bar{\varphi}_{\vk}= \varphi_{- \vk}$ and $\bar{\pi}_{\vk} = \pi_{- \vk}$ (since $(\varphi(\vx), \pi(\vx))$ are real). There is a 1-parameter family of maps ${\cal I}_{\eta}$ from the covariant phase space $\Gamma$ to $\Gamma_{\!\rm can}$: \, $ {\cal I}_{\eta_0} (\phi(\eta,\vx))= (\varphi(\eta_0,\vx), \pi(\eta_0,\vx)) \equiv (\varphi_{\vk} (\eta_0), \bar{\pi}_{\vk}(\eta_0))$, that serves as a symplectomorphism between $\Gamma$ and $\Gamma_{\!\rm can}$ (see, e.g.,  \cite{unitarity}). The metric $g$ on the covariant phase space can be pushed forward by these isomorphisms ${\cal I}_{\eta_{0}}$  to obtain a 1-parameter family of metrics $g_{\rm can}(\eta_{0})$ on $\Gamma_{\!\rm can}$ and now squeezing occurs because the norms that the metrics $g_{\rm can}(\eta_{0})$ assign to the fixed Hamiltonian vector fields $X_{\varphi_{\vk}}$ and $X_{\bar{\pi}_{\vk}}$ change in time $\eta_{0}$. }
\end{enumerate} 

To make these considerations explicit, let us consider three examples that illustrate key aspects of this phenomenon.
\begin{enumerate}
\item  \emph{Minkowski space-time.} This is the simplest example. In this case, the basis functions satisfy
\be
e_k(\eta) = \frac{e^{-ik\eta}}{\sqrt{2k}} \qquad ; \qquad e'_k(\eta) = -i \sqrt{\frac{k}{2}}\, e^{-ik\eta}\, \, ,
\ee
whence (\ref{Eq-Imp1}) and (\ref{Eq-Imp2}) imply
\be \label{gphiphi}
g(\bar{X}_{\varphi_{\vk}(\eta)}, X_{\varphi_{\vk}(\eta)} ) = \frac{1}{k}\, ,
\ee
and
\be \label{gpipi}
g(\bar{X}_{\pi_{\vk}(\eta)}, X_{\pi_{\vk}(\eta)} ) = k \, .
\ee
Thus, in this case the norms of the two Hamiltonian vector fields are \emph{time independent}, reflecting the fact that the dynamical flow now preserves the metric $g$ in addition to the symplectic structure $\Omega$; there is no squeezing. Indeed, a necessary and sufficient condition for the absence of time dependence in quantum uncertainties is that the \emph{dynamical vector field be a Killing field of the metric $g$}. In Minkowski space (and more generally in stationary space-times) it \emph{is} a Killing vector. But in dynamical cosmological space-times, it is not.  Finally, note that the factors of $k$ appear in Eqs (\ref{gphiphi}) and (\ref{gpipi}) only for dimensional reasons, and the product of the norms $||X_{\varphi_{\vk}}||\,||X_{\pi_{\vk}}||$ is equal to 1 at all times.

\item \emph{de Sitter space-time.} Next, let us consider de Sitter space-time. In this case we have
\be
e_k(\eta)= \frac{1}{a(\eta)}\left( 1 + \frac{iHa(\eta)}{k} \right)\,\frac{e^{-ik\eta}}{\sqrt{2k}},\quad {\rm and} \quad e'_k(\eta) = -\frac{i}{a(\eta)}\,\sqrt{\frac{k}{2}}\, e^{-ik\eta}\, .
\ee
Again, it is the Hamiltonian vector fields generated by ${\varphi_{\vk}(\eta)}$ and $\varphi_{\vk'}(\eta)$ are of direct physical interest. Their norms are now given by
\be
g(\bar{X}_{\varphi_{\vk}(\eta)}, X_{\varphi_{\vk}(\eta)} ) = \frac{1}{k}\,\left( \frac{H^2}{k^2} + \frac{1}{a^2(\eta)} \right)\, ,
\ee
and
\be
g(\bar{X}_{\pi_{\vk}(\eta)}, X_{\pi_{\vk}(\eta)} ) = k\, a^2(\eta)\, .
\ee
Thus, not only do norms now change in time, but the effect is rather dramatic in proper time $t$, since $a(\eta)=e^{Ht}$. Let us now examine the evolution of these norms systematically.

Again, the modes of interest are those that are well within the Hubble horizon \emph{at early times} $\eta$, so that $k_{\rm phy} = k/a(\eta) \gg H$. As before, let us suppose that $t=0$ or  $\eta=\eta_0=- (1/H)$ is an early time instant. Since $a(\eta_{0})=1$, modes of interest satisfy $k \gg H$.  At time $\eta_{0}$  the norms of the corresponding HVFs are given by,
\be
||X_{\varphi_{\vk}(\eta_0)}||^2 \approx \frac{1}{k}\qquad {\rm and} \qquad ||X_{\pi_{\vk}(\eta_0)}||^2 = k\, ,
\ee
as in Minkowski space-time.

Let us next consider time $\eta_1$ at which the mode $k$ exits the Hubble horizon; so $k=H a(\eta_1) \equiv - (1/\eta_{1})$. Then,
\be
||X_{\varphi_{\vk}(\eta_1)}||^2\, =\, \frac{1}{k}\, \frac{2H^2}{k^2} \qquad {\rm and} \qquad ||X_{\pi_{\vk}(\eta_1)}||^2\, =\, \frac{k^3}{H^2}\, .
\ee
Since by assumption $k \gg H$ we see that the norm of the HVF associated to the field has decreased while that of the canonically conjugate momentum has increased, both  by a large factor: $||X_{\varphi_{\vk}(\eta_1)}|| \ll ||X_{\varphi_{\vk}(\eta_0)}||$ and $||X_{\pi_{\vk}(\eta_1)}|| \gg ||X_{\pi_{\vk}(\eta_0)}||$. However,  the product of the norms --\,$||X_{\varphi_{\vk}(\eta_1)}||\, ||X_{\pi_{\vk}(\eta_1)}||$\,-- and thus the product of quantum uncertainties, has \emph{only increased by a factor of $2$}.

Finally, let us consider a time $\eta_2$, $N$ e-folds after the mode exits the Hubble horizon, so $\eta_1= e^N \eta_2$. Then, for $N \gg 1$ we have,
\be
||X_{\varphi_{\vk}(\eta_2)}||^2 \approx \frac{H^2}{k^3}\qquad ; \qquad ||X_{\pi_{\vk}(\eta_2)}||^2 = \frac{k^3}{H^2}\,e^{2N}\, .
\ee
Thus, at a late time, the norm $||X_{\varphi_{\vk}(\eta_2)}||$ shrinks only by a factor of $2$ relative to that at horizon crossing, while that of $X_{\pi_{\vk}(\eta_2)}$ has grown exponentially in the number of e-folds.

Let us summarize the situation for the de Sitter space-time. There is enormous squeezing of the norm of the HVF $X_{\varphi_{\vk}(\eta)}$ --mirrored in the uncertainty in $\hat{\varphi}_{\vk}(\eta)$-- between the initial time $\eta_0$, when the norm is essentially the same as in Minkowski space-time, and the time $\eta_1$ at which the mode exits the Hubble horizon. Surprisingly,  there is very little additional squeezing after $\eta=\eta_1$; even if one waits for infinite proper time, the norm approaches a finite value and would have been squeezed only by a factor of 2.  On the other hand, for the momentum $\hat{\pi}_{\vk}(\eta)$  the stretching of the norm of $X_{\pi_{\vk}(\eta)}$ --and hence the uncertainty in $\hat{\pi}_{\vk}(\eta)$-- is quite different. While there is significant stretching between the initial time $\eta_0$ and the horizon crossing time $\eta_1$, the stretching continues to grow exponentially with the number of e-folds after $\eta=\eta_1$. \emph{The product of the norms --and hence of  quantum uncertainties-- grows unboundedly to the future. }

{ The situation in the quasi-de Sitter case is very similar, except that, as we noted in Section \ref{s3.2}, now the scale factor has the form $a(\eta) = (-\mathring{H}\eta)^{- (1+\epsilon_{V})}$ for some constant $\mathring{H}$, and, at late times the mode functions  have an additional time dependence through a multiplicative factor of $(-\eta)^{\delta_{V} -2\epsilon_{V}}$.  At early times,  the norms of the HVFs generated by $\varphi_{\vk}$ and $\pi_{\vk}$ are again essentially the same as in Minkowski space-time.  However,  at late times the leading order terms in these norms are modified.  The norm $||X_{\varphi_{\vk}(\eta_0)}||^{2}$ --and hence the details of the time dependence of squeezing-- change in a manner that depends on the sign of $\delta_{V} - 2\epsilon_{V}$, although the squeezing is not affected in any major way. The norm $||X_{\pi_{\vk}(\eta_0)}||^2$ continues to grow exponentially in the number of e-folds; only the pre-factor $2$ in the exponent is slightly altered. (Here $\epsilon_{V}$ and $\delta_{V}$ are the slow-roll parameters of Eq. (\ref{slowroll}).)}

\item \emph{Radiation filled universe}. Let us consider the radiation filled universe  which has special features.  From (\ref{radiationmodes}) we know that the  basis functions are now given by:
\be
e_k(\eta) = \frac{1}{a(\eta)}\, \frac{e^{-ik\eta}}{\sqrt{2k}} \qquad ; \qquad e'_k(\eta) = - \frac{1}{a(\eta)}\left( \frac{1}{\eta} + ik \right)\, \frac{e^{-ik\eta}}{\sqrt{2k}}\, .
\ee
Using these expressions, we can find the norms of the Hamiltonian vector fields:
\be
||X_{\varphi_{\vk}(\eta)}||^2 = \frac{1}{a^2(\eta)} \frac{1}{k} \qquad {\rm and} \qquad
||X_{\pi_{\vk}(\eta)}||^2 = a^2(\eta)\, \frac{1}{k}\, \left( k^2 + \frac{a_0^4}{4a^2(\eta)} \right) \, .
\ee
Now, in the radiation filled universe, the 4 dimensional scalar curvature vanishes, which means the radius of curvature is infinite,  whence none of the modes cross the curvature radius.%
\footnote{ Here, by curvature radius we mean  $\mathfrak{R}_{\rm curv} = 1/\sqrt{R}$ where $R$ is the space-time scalar curvature. The dynamical equation of mode functions $\phi_{\vk}(\eta)$ --and hence of $e_{k}(\eta)$-- is governed by $\mathfrak{R}_{\rm curv}$. Modes with $k_{\rm phy} \,\mathfrak{R}_{\rm curv} \gg 1$ have oscillatory behavior as in Minkowski space-time, while those with $k_{\rm phy} \, \mathfrak{R}_{\rm curv}  \ll  1$ are `frozen'. In de Sitter space-time  $\mathfrak{R}_{\rm curv}$ coincides with the Hubble radius $(1/H)$ but more generally the two are quite different from each other.}
Hence there is no natural analog of time $\eta_1$ in de Sitter space-time.

For the initial time, let us again choose $\eta_0$ such that $a(\eta_0) = 1$, so $\eta_0 = (2/a_0^2)$ and $t_0= 1/(a_0^2)$. We  then obtain,
\be
||X_{\varphi_{\vk}(\eta_0)}||^2 = \frac{1}{k} \qquad {\rm and} \qquad
||X_{\pi_{\vk}(\eta_0)}||^2 = \frac{1}{k}\, \left( k^2 + \frac{a_0^4}{4} \right) \, .
\ee
The norm $||X_{\varphi_{\vk}(\eta_0)}||$ is the same as in Minkowski space-time, while for modes that are of  high frequency in the sense $k \gg a_0^2$,  the norm $||X_{\pi_{\vk}(\eta_0)}||$ is well-approximated by that in Minkowski space-time. Let us now choose a ``late time" $\eta_2$ such that $a(\eta_2)/a(\eta_0)= e^{\tilde{N}}$. Then, at this late time, the norms satisfy,
\be
||X_{\varphi_{\vk}(\eta_2)}||^2 = \frac{e^{-2\tilde{N}}}{k} = e^{-2\tilde{N}}\, ||X_{\varphi_{\vk}(\eta_0)}||^2\, ,
\ee
and
\be
||X_{\pi_{\vk}(\eta_2)}||^2 = \frac{e^{2\tilde{N}}}{k}\, \left( k^2 + \frac{a_0^4}{4\, e^{2\tilde{N}}} \right)\, \approx \,e^{2\tilde{N}} \,\, ||X_{\pi_{\vk}(\eta_0)}|| \, ,
\ee
where the last approximate equality refers to the high frequency modes with $k \gg a_0^2$.
Thus, between the initial and final time, the norm of $X_{\varphi_{\vk}(\eta)}$ is now squeezed exponentially with the number $\tilde{N}$ of e-folds between $\eta_0$ and $\eta_2$, while that of $X_{\pi_{\vk}(\eta)}$ stretches exponentially. While in both de Sitter and the radiation dominated universes there is significant squeezing, there are two important differences:
\begin{enumerate}
\item Whereas at late times in de Sitter space-time $||X_{\varphi_{\vk}(\eta_2)}||^2 \to ({H^2}/{k^3})$ --which is a $k$-dependent non-zero number-- it goes to zero, exponentially in $\tilde{N}$ for any $k$ in the radiation filled universe. Thus, in contrast to de Sitter space-time, the longer one waits in a radiation-filled universe, more peaked the quantum state becomes; there is no saturation.

\item The product of norms --and hence the quantum uncertainties-- grows exponentially with the number of e-foldings in de Sitter, while it rapidly tends to the Minkowski value in the radiation filled universe.  Note that $||X_{\pi_{\vk}(\eta)}||$ grows exponentially with the number of e-foldings in both cases.  However, as we noted above, there is a striking difference in the squeezing of the norm $||X_{\varphi_{\vk}(\eta)}||$. 
\end{enumerate}

Squeezing is stronger than inflation in the sense of (a) above also in dust filled universes, as well as  the 1-parameter family of FLRW space-times, parameterized by $\alpha \in [0,1]$, that interpolate between the radiation and dust filled cases (see Section \ref{s2.2}) . This is because $(e_{k}(\eta)\, \bar{e}_{k}(\eta))$ falls off as $e^{-2\tilde{N}}$ for mode functions $e_{k}(\eta)$ of (\ref{alphamodes}) for any choice of $\alpha$. These examples explicitly show that inflation is not essential for extreme squeezing. We only need an expanding epoch that lasts for many e-folds in any of these universes.  Interestingly, in the mainstream scenarios the number of e-folds in the radiation + dust epoch is typically \emph{larger} than the (observationally relevant) e-folds during inflation (see footnote {\ref{fn1}}).
\end{enumerate}

Let us summarize the main messages of this section. Squeezing is usually discussed in terms of the time evolution of 2-point functions $\langle  \hat{\varphi}^{\dag}_{\vk}(\eta) \,\hvpk(\eta) \rangle$ and $\langle  \hat{\pi}^{\dag}_{\vk}(\eta) \,\hat\pi_{\vk} (\eta) \rangle$ that encode  quantum uncertainties. We saw that  this phenomenon can be traced back to the classical phase-space: it is entirely captured in the time evolution of norms of the vector fields $X_{\vpk(\eta)}$ and $X_{\bar\pi_{\vk} (\eta)}$ on $\Gamma$. The norms themselves are calculated using the metric $g$ --the new geometrical structure on the phase space needed in the passage to quantum theory. Thus, the origin of the squeezing phenomenon can be directly traced back to geometrical structures on the phase space. Since this characterization naturally extends to fields $\phi$ on all FLRW backgrounds, \emph{the phenomenon is not tied to inflation}. The description in terms of the evolution of norms also served to bring out conceptually important subtleties --e.g., most of the squeezing occurs before the mode exits horizon in de Sitter space-time, and there are interesting differences between squeezing in de Sitter and radiation or dust filled space-times.\\

We will conclude this discussion with a few remarks.

1.  Note that the phenomenon of squeezing is tied to the choice of canonically conjugate pair of observables: Given a quantum state and dynamics, one can have squeezing with respect to one set of such observables and no squeezing with respect to another. For example, the canonically conjugate pairs $\hat{q}_{\vk}$ and $\hat{p}_{\vk}$ of Eq. (\ref{ip})  undergo no squeezing. But because these observables are ($\eta$-dependent)  linear combinations of  $\hat{\varphi}_{\vk}(\eta)$\,\emph{and}\, $\hat{\pi}_{\vk}(\eta)$, they are difficult to measure and are not of directly physical interest. What one measures in the CMB is $|\Delta\hat{\varphi}_{\vk}(\eta_{\,\rm ls})|$ through the power spectrum at the time $\eta_{\,\rm ls}$, corresponding to the surface of last scattering. (One does not directly measure $|\Delta\hat{\pi}_{\vk}(\eta_{\,\rm ls})|$ either).

2.  In ordinary quantum mechanical systems, if $|\Delta\hat{p}|$ is large at an instant $\tau$ of time, then typically  this uncertainty spreads to $\hat{q}$ soon there after. We saw that during inflation as well as radiation dominated era,  the uncertainty $|\Delta\hat{\pi}_{\vk}(\eta)|$ increases exponentially with the number of e-folds, while the uncertainty $|\Delta\hat{\varphi}_{\vk}(\eta)|$ either tends to a constant (de Sitter) or even decreases exponentially (radiation or dust-filled universe). So there seems to be  an apparent paradox. To  see why there is no contradiction, let us recall our discussion of the grandfather clock in Section \ref{s1}  (or consider any macroscopic system with a large mass $M$). In that case, the uncertainty $|\Delta \hat{p}|$ in momentum translates into an uncertainty $(|\Delta \hat{p}|\,/M)$ in velocity that then descends to the uncertainty $|\Delta \hat{x}|$ in position later on. As explicit numbers showed in Section \ref{s1}, $|\Delta \hat{x}|$ can remain small  for a very long time if  the mass $M$ is sufficiently large. Similarly,  since $a(\eta)$ is so large at late times and $\vpk^{\prime}(\eta) = \pi_{\vk}(\eta) / a^{2}(\eta) $,\,\,$|\Delta \hvpk |$ continues to remain small in the distant future.

3.  As in Section \ref{s3}, we focused our attention on just two modes, $\vk$ and $-\vk$, also in the discussion of squeezing. However,  again the reasoning can be extended to the full Klein-Gordon field $\hvp (\vx, \eta)$ and its conjugate momentum $\hat\pi(\vx, \eta)$ by restricting oneself to suitable \emph{subset} of configuration and momentum observables $(\hat\varphi(f)) (\eta)$  and $(\hat\pi(g)) (\eta)$, obtained by smearing $\hvp (\vx, \eta)$ and  $\hat\pi(\vx, \eta)$ with appropriate test functions $f(\vx)$ and $g(\vx)$. If these test functions are chosen so that their Fourier transforms have support only on a finite but arbitrarily large band of modes, then the norms of the Hamiltonian vector fields of the phase space functions $(\varphi(f))(\eta)$ and $(\pi(g)) (\eta)$ exhibit the same squeezing and stretching behavior. And again this behavior translates directly to quantum uncertainties in $|\Delta \varphi(f)|$ and $|\Delta\pi(g)|$.  { Finally, while we used specific vacua in the illustrative space-times considered, the entire discussion goes through for any homogeneous isotropic quasi-free quantum state in any FLRW space-time: Each of these states arises from a K\"ahler structure on $\Gamma$, and its 2-point function is determined by the corresponding positive definite metric $g$.}

\section{Approximating the quantum state by a phase space probability distribution function}
\label{s5}

Recall that in post-inflationary dynamics, the quantum state of cosmological perturbations is generally replaced by an appropriate mixed classical state --i.e. a distribution function on the classical phase space. All subsequent analysis is then classical.  Can this procedure be justified from first principles? Since quantum theory has a much richer content than classical, does this procedure not throw out, by fiat, some essential aspect of quantum non-commutativity and quantum dynamics that may be observationally relevant? In this section we will systematically analyze this issue.
 
The analysis becomes most transparent in the Bargmann representation \cite{bargmann1, bargmann2} because quantum states are now represented as (holomorphic) functions on the \emph{phase space}: the relation between quantum and classical structures is brought to the forefront,  enabling us to obtain a sharp result that, to our knowledge, has not been discussed in the literature. { Generally calculations have been carried out in  the configuration representation in which states are functions only of the configuration variables  --and sometimes in the momentum representation, where states are functions only of momenta-- rather than on the full phase space.}

Let us consider \emph{any} FLRW background. Given a classical observable $\mathcal{O}$ constructed from \emph{arbitrary} sums of products of the canonically conjugate pairs  $\vp_{\vk}(\eta)$ and $\bar\pi_{\vk}(\eta)$,  let $\hat{\mathcal{O}}_{W}$ be the corresponding Weyl-ordered quantum operator.%
\footnote{ $\hat{\mathcal{O}}_{W}$ is the totally symmetric self-adjoint operator corresponding to the classical observable ${\mathcal{O}}$. Explicit definition is given below  Eq.(\ref{key1}).}
Consider the Fock representation selected by a basis $e_{k}(\eta)$ (that defines a complex structure $J$ compatible with the symplectic structure $\Omega$).  The Bargmann representation naturally associates with the Fock vacuum  $\Psi_{o}$ a probability distribution function $\rho_{o}$ on the phase space $\Gamma$,  using the metric $g$ thereon  (defined by $J$ and $\Omega$; see Eq. (\ref{metric1})). 
We will show that \emph{the quantum expectation value $\langle\hat{\mathcal{O}}_{W} \rangle_{{}_{0}}$ of $\hat{\mathcal{O}}_{W}$ in the state $\Psi_{o}$ exactly equals the classical expectation value $\langle \mathcal{O} \rangle_{\rho_{o}}$ for all $\mathcal{O}$.}  This seems surprising at first since the family 
of these $\mathcal{O}$ is so large as to contain every observable that is generally considered. Where does the richer information in the quantum theory then reside? It resides in the expectation values of sums of products of quantum operators that are \emph{not} Weyl ordered.  More precisely, any operator $\hat{\mathcal{O}}$ can be brought to its Weyl ordered form by performing permutations. But because of the quantum non-commutativity, these permutations generate additional terms that vanish in the $\hbar \to 0$ limit. The richer information that eludes the classical theory is succinctly captured in the expectation values of these additional terms that have to be included for operators that are not Weyl ordered. 

The streamlined procedure provided by the Bargmann representation and the final result  is quite general. In the particular application to cosmological perturbations, it provides a clearcut  justification for the procedure used in the early universe literature (see also, e.g., \cite{dpaas,lps,kiefer1,kiefer2,jmvv1} for other lines of reasoning). We will divide our discussion into two steps because one encounters certain technical complications in the application to cosmological perturbations. While these complications are not significant conceptually, their presence  can obscure the underlying ideas. Therefore, in Section \ref{s5.1} we will present the crux of the reasoning using a simple quantum mechanical system and then apply the ideas to the cosmological setting in Section \ref{s5.2}.

\subsection{Bargmann representation: Relation between quantum and classical structures}
\label{s5.1}

Let us begin with the simplest context: quantum mechanics of a particle in 1 spatial dimension. As is common in textbooks,  let us remove the inessential numerical and dimensional factors and consider operators $\hx$ and $\hp$ satisfying the commutation relations $[\hx ,\, \hp] = i$. Consider the  usual annihilation and creation operators  $\ha = (\hx + i\hp)/\sqrt{2}$, and $\had = (\hx - i\hp)/\sqrt{2}$ satisfying $[\ha, \, \had ] =1$.  In the Bargmann representation \cite{bargmann1} the abstract algebra generated by $\hx, \, \hp$ is represented by concrete  operators on a Hilbert space of functions on the \emph{phase space} $\Gamma$.  Let us set
\be \label{bargmeasure}  \zeta = x + ip, \qquad {\rm and} \qquad \rmd\mu_{\rm B} \,=\,\f{1}{\pi} \,e^{-\zeta\bar\zeta}\,\, \rmd \mu_{\rm L} \,\equiv\, \f{1}{\pi} e^{-\zeta\bar\zeta}\,\,\rmd x \rmd p \, ,\ee
where $\rmd\mu_{\rm B}$ and $\rmd\mu_{\rm L}$ are, respectively, the Bargmann and Liouville measures on the phase space.  While the Liouville volume of phase space is infinite, the Bargmann measure is normalized: $\int_{\Gamma} \rmd \mu_{\rm B} = 1$. { In terms of geometric structures introduced in Section \ref{s2} and used in the discussion of squeezing in Section \ref{s4}, the exponent in the Bargmann measure is just the norm $g(\bar\zeta,\, \zeta)$ of the complex vector $\zeta$ in $\Gamma$, defined by the metric $g$ that dictates squeezing.} 

Quantum states are entire holomorphic functions $\Psi(\zeta)$ on  $\Gamma$ equipped with the inner product
\be\label{bargip} \langle \Phi,\, \Psi \rangle = \int_{\Gamma} \overline{\Phi (\zeta)}\, \Psi(\zeta)\,\, \rmd \mu_{\rm B}\, . \ee
Note that while each $\Psi(\zeta)$ diverges at infinity, the norm is still finite for any polynomial $\Psi(\zeta) = \sum_{n=1}^{N} \Psi_{n}\,\zeta^{n}$ because of the exponential damping $e^{-\zeta\bar\zeta}$ in the Bargmann measure. Also because of this factor,  and because $\bar\Phi$ is anti-holomorphic in $\zeta$, we can represent $\ha$ and $\had$ as
\be \label{bargrep} \had \Psi(\zeta) = \zeta \Psi(\zeta)\qquad {\rm and} \qquad \ha \Psi(\zeta) = \f{\rmd \Psi}{\rmd \zeta} (\zeta)\, ; \ee
these concrete operators satisfy the desired commutation relations. It follows from the action (\ref{bargrep}) of creation and annihilation operators that this  representation of the canonical commutation relation is irreducible and hence unitarily equivalent to the more familiar position (or  momentum) representation in which quantum states arise as square-integrable functions only of $x$ (or, $p$).

Let us note a few features of the Bargmann representation that will be directly useful in what follows. First, it follows from the definition (\ref{bargrep}) of the  annihilation operator $\ha$ that $\Psi_{\zeta_{o}} := e^{\zeta_{o}\, \zeta}$ is a normalizable eigenstate of $\ha$ with the following properties:
\be  \ha\Psi_{\zeta_{o}} = \zeta_{o} \Psi_{\zeta_{o}}; \qquad {\rm whence} \qquad \langle \Psi_{\zeta_{o}},\, (\hx+ i\hp) \Psi_{\zeta_{o}}\rangle = \sqrt{2} \zeta_{o}\, ,\ee 
so that $\Psi_{\zeta_{o}}$ is a coherent state peaked at the phase space point $\zeta= \sqrt{2} \zeta_{o}$, and, 
\be \big(\Delta \hx \big)^{2}_{\Psi_{\zeta_{o}}} =  \big(\Delta \hp \big)^{2}_{\Psi_{\zeta_{o}}} = \f{1}{2}\, , \ee
whence the uncertainty on $\hx$ and $\hp$ are equally distributed and the product of these uncertainties is saturated. In particular, $\Psi_{o}(\zeta) =1$ is the coherent state peaked at $ x=0, p=0$; this is the (normalized) vacuum state.  Finally, note that the monomials:
\be \label{bargbasis} \Psi_{n}(\zeta)\,  :=\, \f{ \zeta^{n}}{\sqrt{n!}} \ee
provide an orthonormal basis in the Bargmann Hilbert space. These correspond to the $n$th excited states of a harmonic oscillator, although the Bargmann representation by itself is only a kinematical construct that does not refer to any specific Hamiltonian. (We note in passing that the squeezed states considered in Section \ref{s4} also have a simple representation: $\Psi (\zeta) = e^{\mathfrak{s}\, \zeta^{2}}$ where $\mathfrak{s}$ is a complex constant that characterizes squeezing in $\hx$ and $\hp$.)

Of particular interest is the 2-parameter family of Weyl operators,%
\footnote{\label{fn10}The Weyl operators are of special interest to Fock quantization. It is easy to verify that they satisfy: $\hat{W}(\lambda_{1},\,\mu_{1})\, \hat{W}(\lambda_{2},\,\mu_{2})  = e^{-\f{i}{2} (\lambda_{1} \mu_{2} - \lambda_{2}\mu_{1})}\, \hat{W} (\lambda_{1}+\lambda_{2},\, \mu_{1}+ \mu_{2})$. Therefore, the vector space generated by linear combinations of the $\hat{W}(\lambda,\,\mu)$ is closed under products; it is an algebra.  This is the Weyl algebra $\mathfrak{W}$ that has the full information-content of  the standard Heisenberg algebra generated by the canonically conjugate operators. The Fock representation is completely characterized by the vacuum expectation value $\langle \hat{W}(\lambda,\,\mu)  \rangle_{{}_{0}} $ of Weyl operators; as noted below, all $n$-point functions can be obtained by taking derivatives of this vacuum expectation value with respect to $\lambda$ and $\mu$.}
\be \hat{W}(\lambda,\,\mu)\,  := \,e^{i\,(\lambda \hx+ \mu\hp)} \,\,{\equiv\, e^{-\f{1}{2} \bar\alpha\alpha} \, e^{-\bar\alpha \had}\, e^{\alpha\ha} }\ee
where $\lambda, \mu$ are real and $\alpha = (\mu+i\lambda)/\sqrt{2}$. In the Bargmann representation their action is given by:
\be \hat{W}(\lambda,\,\mu)\, \Psi(\zeta) = e^{-\f{1}{2} \alpha\bar\alpha}\, e^{-\bar\alpha \zeta} \, \Psi (\zeta+\alpha)\, .   \ee
To bring out the relation between expectation values of quantum and classical observables, let us begin by considering  the vacuum expectation values (VEVs) of Weyl operators:
\ba \label{weylvev}  
\langle \hat{W}(\lambda,\,\mu)  \rangle_{{}_{o}} &:=& \langle \Psi_{o},\,  e^{i(\lambda \hx + \mu\hp)} \Psi_{o} \rangle \,\, = \langle \Psi_{o},\,   e^{\alpha \ha - \bar\alpha \had} \, \Psi_{o}\rangle \nonumber \\
&=&  e^{-\f{1}{2} \alpha\bar\alpha}\, \langle\Psi_{o} ,\,   e^{-\bar\alpha \had}\, e^{\alpha \ha}\, \Psi_{o} \rangle \,\,
=e^{-\f{1}{4}\, (\lambda^{2}+\mu^{2})} \ea
where, in the last step we used the fact that $\Psi_{o}$ is an eigenstate of $e^{\alpha \ha}$ with eigenvalue $1$ since the vacuum is represented by the Bargmann state $\Psi_{o}(\zeta) =1$.
It is clear from the very definition of $\hat{W}(\lambda,\mu)$  that successive derivatives of (\ref{weylvev})  with respect to $\lambda$ and $\mu$ provide VEVs of products of $\hx$ and $\hp$. More precisely,  it is straightforward to verify that 
\be \label{key1} 
\f{1}{i^{m+n}}\,\,{ \f{\partial ^{n}}{\partial \lambda^{n}}\,  \f{\partial ^{m}}{\partial\mu^{m}}}\, \langle \hat{W}(\lambda,\,\mu)  \rangle_{{}_0} \,\Big{|}_{\lambda=0, \mu=0}\!= \,\langle\,  \big(\hx^{n} \hp^{m}\big)_{W} \rangle_{{}_{0}}\, , \ee
where the Weyl order product $\big(\hx^{n} \hp^{m}\big)_{W} $ is the sum of all\,  ${(m+n)!}/{(m!n!)}$\, permutations of the product $\hx^{n}\hp^{m}$ --each of which contains $n$ operators $\hx$ and $m$ operators $\hp$-- divided by the number of terms. (Thus, for example, $(\hx\hp)_{{}_{W}} = \f{1}{2} \big(\hx\hp + \hp\hx)$;\,\, $(\hx^{2}\hp)_{{}_{W}} = \f{1}{3} \big( \hx^{2}\hp+ \hx\hp\hx + \hp\hx^{2}\big)$; etc.) 
\medskip

{ Let us now turn to the question of finding a classical mixed state that is to correspond to the quantum vacuum $\Psi_{o}$. Since  $\Psi_{o}(\zeta) =1$,  the probability distribution $\rho_{o}$  it defines on $\Gamma$ (with respect to the Liouville measure $\rmd\mu_{\rm L}$) is given by}
%
\be \label{corres1} \rho_{o} =  \f{1}{\pi} \,\,e^{-\zeta\bar\zeta}  \equiv \f{1}{\pi}\,\, e^{-(x^{2}+p^{2})} \qquad {\rm with} \qquad \int_{\Gamma} \rho_{o} \rmd\mu_{\rm L} =1\, .  \ee
So it is natural to consider $\rho_{o}$ as the distribution function on $\Gamma$ that corresponds to the quantum vacuum $\Psi_{o}$. Next, as we saw, the family of Weyl operators provides a convenient tool to compute expectation values of polynomials in the basic canonical variables. Let us then consider the classical analogs 
\be {W}(\lambda,\,\mu) := e^{i\,(\lambda x\,+\, \mu p)} \ee
of Weyl operators and calculate their expectation values in the distribution function $\rho_{o}$. We have:
\ba \label{weylvevcl}
\langle W(\lambda,\mu) \rangle_{\rho_{o}}  &:=& \int_{\Gamma} W(\lambda,\mu)\,  \rho_{o}\, \rmd \mu_{\rm L} = \f{1}{\pi}\,\int_{\Gamma}  e^{i\,(\lambda x\,+\, \mu p)} \, e^{-(x^{2}+p^{2})} \,\rmd x \rmd p\nonumber\\
&=& e^{-\f{1}{4}\, (\lambda^{2}+\mu^{2})} \,\equiv\, \langle \hat{W}(\lambda,\,\mu)  \rangle_{{}_{0}} \, .\ea
%
Thus, the quantum and classical expectation values of all Weyl observables are equal!  This exact equality is striking, especially because the quantum vacuum is completely determined by the VEVs of the Weyl operators (e.g. through the Gel'fand-Naimark-Segal construction \cite{gns1,gns2}; see also footnote \ref{fn10}).  In particular, we can take repeated derivatives of $\langle W(\lambda,\mu) \rangle_{\rho_{o}}$ with respect to $\lambda$ and $\mu$ and evaluate the result at $\lambda=0,\, \mu=0$ to obtain the expectation values of arbitrary polynomials in $x$ and $p$:
\be \label{key2} 
\f{1}{i^{m+n}}\,\,{ \f{\partial ^{n}}{\partial \lambda^{n}}\,  \f{\partial^{m}}{\partial \mu^{m}}}\, \langle {W}(\lambda,\,\mu)  \rangle_{\rho_{o}}\,\Big{|}_{\lambda=0, \mu=0}\!= \,\langle\,  \big(\hx^{n} (\hp)^{m}\big) \rangle_{\rho_{o}}\, , \ee
Therefore the equality $\langle \hat{W}(\lambda,\,\mu)  \rangle_{{}_{0}} = \langle W(\lambda,\mu) \rangle_{\rho_{o}}$  implies:
\be \label{key3}
  \langle\,  \big(\hx^{n} \hp^{m}\big)_{W} \rangle_{{}_{0}}\,  = \,  
  \langle\,  x^{n} p^{m} \rangle_{\rho_{o}}\, , \ee
for all $m,n$. 
Thus the expectation value of an \emph{arbitrary} monomial in $x$ and $p$ in the classical state $\rho_{o}$ \emph{is exactly equal to} the expectation value in the quantum state $\Psi_{o}$ of the \emph{Weyl ordered version} of that monomial in $\hx$ and $\hp$. Put differently, consider \emph{any} polynomial $F(x,p)$ in $x$ and $p$ on the classical phase space. Then there is an explicit factor ordering procedure that yields a quantum operator $\hat{F}$ such that the expectation value of that $\hat{F}$ in the quantum state $\Psi_{o}$ is the same as the expectation value of $F$ in the classical state $\rho_{o}$. The correspondence $F \leftrightarrow \hat{F}$ is both unique and explicit. Again, this equality is striking because of the underlying universality: it holds for \emph{any}  $F(x,p)$.

Next, recall that $\Psi_{o}$ is a coherent state peaked at $x=0, p=0$.  It is therefore natural to ask if the correspondence $\Psi_{o} \leftrightarrow \rho_{o}$ in the strong sense of Eq. (\ref{key3}) extends to all coherent states. The answer is in the affirmative. Recall that in the Bargmann representation a general normalized coherent state  peaked at the phase space point $\zeta \equiv x+ip = \sqrt{2} \zeta_{o}$ is represented by the holomorphic function 
$\Psi_{\zeta_{o}} = e^{-\f{1}{2} \bar\zeta_{o}\zeta_{o}}\, e^{\zeta_{o}\zeta}$ on $\Gamma$. Therefore, the same reasoning that led us to Eq. (\ref{corres1}) now suggests that we set the correspondence 
\be  \Psi_{{\zeta_{o}}} = e^{-\f{1}{2} \bar\zeta_{o}\zeta_{o}}\, e^{\zeta_{o}\zeta} \qquad \leftrightarrow \qquad \rho_{{}_{\zeta_{o}}} = \f{1}{\pi} \,e^{-(\zeta - \zeta_{o})(\bar\zeta - \bar\zeta_{o})}\, .  \ee
Then one can verify that the direct analog of (\ref{key3})
\be \label{coherent} \langle\,  \big(\hx^{n} \hp^{m}\big)_{W} \rangle_{{}_{{\zeta}_{o}}}\,  = \,  
  \langle\,  x^{n} p^{m} \rangle_{{}_{\rho_{{}_{\zeta_{o}}}} }\, , \ee
holds, again establishing the emergence of classical behavior for a general coherent state $\Psi_{\zeta_{o}}$. It is natural to ask whether this result can be extended to superpositions of coherent states. More precisely, let $\Psi_{\zeta_{1}}$ and $\Psi_{\zeta_{2}}$ be two (normalized) coherent states and $\rho_{{}_{\zeta_{1}}}$ and $\rho_{{}_{\zeta_{2}}}$ the corresponding classical distribution functions. Consider the normalized superposition  $\Psi = c_{1} \Psi_{\zeta_{1}} + c_{2} \Psi_{\zeta_{2}}$ with $|c_{1}|^{2} + |c_{2}|^{2} =1$. Would the equality (\ref{coherent}) continue to hold if we replace $\Psi_{\zeta_{o}}$ on the left side with $\Psi$ and $\rho_{o}$ on the right side with the normalized probability distribution function  $\rho = |c_{1}|^{2} \rho_{{}_{\zeta_{1}}} + |c_{2}|^{2} \rho_{{}_{\zeta_{2}}}$? The answer is in the negative because the left side in Eq. (\ref{coherent})  is quadratic in the state, whence it would now include `cross-terms' which do not vanish because coherent states are not orthogonal to each other.
\smallskip

Finally, let us note the generality of all these considerations. Suppose we make a linear canonical transformation on the phase space to pass to new canonical coordinates $(x^{\prime}, \, p^{\prime})$. Then we can define $\zeta^{\prime}$ via $\zeta^{\prime} = x^{\prime} + i p^{\prime}$ and construct the Bargmann representation using $\zeta^{\prime}$ (which is unitarily equivalent to the original unprimed representation). We can again associate with the new quantum vacuum $\Psi_{o}^{\prime}$ a classical distribution function $\rho^{\prime}_{o}$ and show that the  primed analog of (\ref{key3}) holds. Now, in the original unprimed representation, the state  $\Psi_{o}^{\prime}$ is represented by a \emph{squeezed state} in which it is the uncertainties in $\hx^{\prime}$ and $\hp^{\prime}$ that are equally distributed (and their product is again minimized). Thus, the equality of expectation values 
is not limited to coherent states. Given \emph{any squeezed state} $\Psi_{o}^{\prime}$ in the original Bargmann representation, we can also find a distribution function $\rho_{o}^{\prime}$ on the classical phase space $\Gamma$ such that there is an exact equality  between the expectation values of polynomial classical observables in the classical state $\rho^{\prime}_{o}$  and those of specific quantum lifts of these observables in the quantum state $\Psi_{o}^{\prime}$.

This rich interplay between quantum and classical states is brought to the forefront in a streamlined fashion in the Bargmann representation because now quantum states are represented as functions on the \emph{full phase space}; one does not break the symmetry between configuration and momentum variables by going to the $x$ or the $p$ representation.\\

\emph{Remarks:}  

{ 1. For simplicity of presentation we worked with dimensionless variables $x, p, \lambda,  \mu$. To restore the correct physical dimensions, one can introduce a length scale $d$ and define tilde variables $\tilde{x} = xd,\, \tilde{p} = \hbar{p}/d,\, \t\lambda = d \lambda, \t\mu= (d/\hbar)\mu$  (keeping  $\hat{\tilde{a}} = \ha,\, \t\zeta = \zeta, \, \rmd\tilde{\mu}_{\rm L} = \rmd\mu_{\rm L}, \, \rmd\tilde{\mu}_{\rm B} = \rmd\mu_{\rm B}$). Then the Bargmann representation and Weyl operators remain unchanged and we can simply recast the final result (\ref{key3}) in terms of physical canonical variables $\tilde{x}, \, \tilde{p}$: 
\be  \label{key3tilde} 
\langle\,  \big(\hat{\tilde{x}}^{n} \hat{\tilde{p}}^{m}\big)_{W} \rangle_{{}_{0}}\,  = \,   \langle\,  \tilde{x}^{n} \tilde{p}^{m} \rangle_{\rho_{o}}\,. \ee
A small subtlety is that, when expressed in terms of physical variables $\tilde{x},\,\tilde{p}$,  the distribution function on the classical phase space becomes $\rho_{o} = \exp -\big((x/d)^{2} + (pd/\hbar)^{2}\big)$; it knows about $\hbar$.  This is inevitable because the VEV on the left side of  (\ref{key3tilde})  knows about $\hbar$, while the term $\tilde{x}^{n}\tilde{p}^{m}$ on the right side doesn't.

2. In the cosmology literature, main emphasis has been on the configuration representation, where quantum states are square-integrable functions of $x$.  In this representation,  it is very easy to establish the correspondence between quantum and classical expectation values of observables $f(x)$ that depend only on position; and indeed one can do that for any normalized quantum state $\psi(x)$: associate with $\psi(x)$ the distribution function $\rho = |\psi(x)|^{2} \sigma(p)$, where $\sigma(p)$ is \emph{any} normalized positive function of $p$, to trivially obtain $ \langle (f(\hat{x}) \rangle_{\psi}\,  = \,   \langle\, f(x) \rangle_{\rho}$. But in contrast to the Bargmann representation,  that treats $x$ and $p$ on equal footing,  it is cumbersome to treat general polynomials $F(x,p)$ that depend both on position and momenta. 


{ 3. For simplicity, in this subsection we did not explicitly consider time evolution. If the equations of motion are linear, then the time dependence in $x(t), p(t)$ naturally lifts to the quantum theory and provides us with $\hat{x}(t), \hat{p}(t)$. One can work with time dependent Weyl operators $(\hat{W}(\lambda,\mu) (t)) :=  e^{i\,(\lambda \hx(t)+ \mu\hp(t) )}$ and their classical counterparts. Because the evolution preserves the symplectic structure, the entire discussion goes through and we obtain the equality  $\langle\,  \big(\hx^{n}(t) \hp^{m}(t)\big)_{W} \rangle_{{}_{0}}\,  = \,  \langle\,  x^{n} (t)p^{m}(t) \rangle_{\rho_{o}}$ for all $t$.  This time dependence is included explicitly in the application of the Bargmann representation to cosmology that follows.}

\subsection{Application to cosmology}
\label{s5.2}

Let us now return to quantum fields $\hat\phi (\vx, \eta)$ on a general FLRW space-time. We will now show that one can again construct the Bargmann representation and arrive at the analog of the exact equality (\ref{key3}), thereby establishing the desired emergence of classical behavior. However, there are three technical points we have to take into account in order to repeat the procedure used in the last subsection:\\
(i) unlike the canonical variables $(x,\, p)$ of Section \ref{s5.1}, now  the basic canonically conjugate variables  $(\vp_{\vk},\, \pi_{-\vk})$ are complex-valued;\\
(ii)  they are not independent, since $\bar\vp_{\vk}(\eta) = \vp_{-\vk}(\eta)$ and $\bar\pi_{\vk}(\eta) = \pi_{-\vk}(\eta)$; and,\\ 
(iii)  the  momentum conjugate to $\vp_{\vk}$ is $\pi_{-\vk}$ rather than $\pi_{\vk}$.  While these technicalities make intermediate expressions more complicated, they do not affect the conceptual reasoning nor the underlying mathematical structure. 

Because of point (iii) above, now we are forced to consider two degrees of freedom  --\,modes $\vk$ and $-\vk$\,-- whence the phase space of interest will be 4 (real) dimensional. Let us simplify the notation by setting:
\ba \hvp_{1}(\eta) &:=&\f{1}{\sqrt{\hbar}}\, \hvpk(\eta), \qquad  \hat\pi_{1}(\eta) :=\f{1}{\sqrt{\hbar}}\, \hat\pi_{-\vk}(\eta), \nonumber\\
\hvp_{2} (\eta) &:=& \f{1}{\sqrt{\hbar}}\, \hvp_{-\vk}(\eta), \qquad \hat\pi_{2}(\eta) := \f{1}{\sqrt{\hbar}} \,\hat\pi_{\vk}(\eta) \ea
%
%
so that $[\hvp_{I}(\eta), \, \hat\pi_{J}(\eta)] = i \delta_{IJ}$, \, where $I, J = 1,2$,\,\, and also define  
\be \ha_{1} = \f{1}{\sqrt\hbar}\, \hat{A}_{\vk}\qquad {\rm and}\qquad  \ha_{2} = \f{1}{\sqrt\hbar}\, \hat{A}_{-\vk} \ee
so that $[\ha_{I},\, \had_{J}] = \delta_{IJ}$.  Next,  setting $e(\eta) = e_{k}(\eta)$ we now have 
\ba \hvp_{1}(\eta)  &=& e(\eta) \ha_{1} + \bar{e}(\eta) \had_{2} \qquad \hat\pi_{1} = a^{2}(\eta) \big(\bar{e}^{\prime}(\eta) \had_{1} + e^{\prime}(\eta) \ha_{2} \big)\nonumber\\
\hvp_{2}(\eta)  &=& e(\eta) \ha_{2} + \bar{e}(\eta) \had_{1} \qquad \hat\pi_{2} = a^{2}(\eta) \big(\bar{e}^{\prime}(\eta) \had_{2} + e^{\prime}(\eta) \ha_{1} \big)\, . \ea
Given any complex numbers  $\lambda, \mu$, the combinations $\lambda \hvp_{1}(\eta)\, +\, \bar\lambda \hvp_{2}$ and $\bar\mu\hat\pi_{1}(\eta)\,  +\, \mu \hat\pi_{2}(\eta)$ are both self-adjoint since $\hvp_{1}^{\dag} = \hvp_{2}$ and $\hat\pi_{1}^{\dag} = \hat\pi_{2}$. Therefore,  we can now define the Weyl  operators:
\be \label{weyl} \hat{W}(\lambda, \mu) (\eta) := e^{i\big( \lambda \hvp_{1}(\eta)\, +\, \bar\lambda \hvp_{2}(\eta) \,+\, \bar\mu\hat\pi_{1}(\eta)\,  +\, \mu \hat\pi_{2}(\eta) \big)} = {e^{\, \sum_{I} \, \alpha_{I}(\eta) \ha_{I} -\bar\alpha_{I}(\eta) \had_{I} \, }} \ee
where we have set 
\be { \alpha_{1}(\eta) = i\big(\lambda\, e(\eta) + \mu a^{2}(\eta) \,e^{\prime} (\eta)\big)\quad {\rm and} \quad 
\alpha_{2}(\eta) = i\big(\bar\lambda\, e(\eta) + \bar\mu a^{2}(\eta)\, e^{\prime} (\eta)\big) \, .}\ee
Next, as a prelude to the Bargmann representation, let us introduce the complex coordinates $\zeta_{I}$  on phase space simply by rescaling, for later convenience,  the coordinates $z_{\vk}$ introduced in  Eq.  (\ref{expfield}) : 
\be {\zeta_{1} = \sqrt{{2}/{\hbar}}\,\, z_{\vk} \qquad {\rm and}\qquad \zeta_{2} = \sqrt{{2}/{\hbar}} \, \, z_{-\vk}}\, .\ee 
Since there are no relations between $z_{\vk}$ and $z_{-\vk}$,\, $\zeta_{I}$ are independent and coordinatize the 4-dimensional phase space; they are the analogs of  (the dimensionless) $\zeta= x+ ip$ used in Section \ref{s5.1}.  (See also the remark on real phase space coordinates $q_{\vk}, p_{\vk}$ in Section \ref{s4.2}.)

Quantum states in the Bargmann representation are then phase space functions $\Psi(\zeta)$ that are holomorphic in $\zeta_{I}$, endowed with the inner product
\be \label{ipBargmann} 
\langle \Phi, \Psi \rangle = \int \overline\Phi \, \Psi\,\, \rmd\mu_{\rm B}  \equiv \int \overline\Phi \, \Psi \,\,e^{-\sum_{I} \bar\zeta_{I} \zeta_{I}}\,\,\rmd\mu_{\rm L}\, ,
\ee
where $\rmd\mu_{\rm L}$ is again the Liouville measure. The creation and annihilation operators are represented by
\be \had_{I} \Psi (\zeta) = \zeta_{I} \Psi(\zeta) \qquad {\rm and} \qquad \ha_{I} \Psi (\zeta) = \f {\partial\Psi}{\partial \zeta_{I}} (\zeta) \ee
and the vacuum state $\Psi_{o}(\zeta)$, annihilated by the two $\ha_{I}$, is again given by $\Psi_{o}(\zeta) =1$.
Therefore, in view of (\ref{weyl}), the VEV of the Weyl operators is given by
\ba \langle \hat{W}(\lambda,\,\mu)(\eta)  \rangle_{{}_{o}} &:=& \langle \Psi_{o},\,\, {e^{\sum_{I} \, \alpha_{I}(\eta) \ha_{I} \, - \bar\alpha_{I}(\eta) \had_{I} } }\,\Psi_{o}\rangle\nonumber\\
&=& e^{-\f{1}{2} \sum_{I} \alpha_{I}(\eta) \bar\alpha_{I}(\eta)}\,\, \langle \Psi_{o},\,\, { e^{-\sum_{I} \bar\alpha_{I} (\eta)\had_{I}}\,\, e^{\sum_{I} \alpha_{I}(\eta) \ha_{I}}\, }\Psi_{o} \rangle\nonumber\\
&=& e^{-\f{1}{2} \sum_{I} \alpha_{I}(\eta) \bar\alpha_{I}(\eta)}\ea
where, in the last step, we have used the fact that $e^{\beta_{I} \ha_{I}} \Psi_{o} = \Psi_{o}$ for arbitrary complex numbers $\beta_{I}$. 

Let us now turn to the classical theory. Considerations of Section \ref{s5.1} suggest that the quantum vacuum $\Psi_{o}$ should be matched with  the distribution function $\rho_{o} =\f{1}{\pi^{2}}\, e^{-\sum_{I} \zeta_{I}(\eta) \bar\zeta_{I}(\eta)}$ on the classical phase space. Therefore, the classical expectation values of the Weyl functions are given by
\ba \langle W(\lambda,\,\mu)(\eta) \rangle_{\rho_{o}} &=& \int\,
e^{i\big( \lambda \vp_{1}(\eta)\, +\, \bar\lambda \vp_{2}(\eta) \,+\, \bar\mu\pi_{1}(\eta)\,  +\, \mu \pi_{2}(\eta) \big)}\,\, \underbrace{\f{1}{\pi^{2}}\,e^{-\sum_{I} \zeta_{I}(\eta) \bar\zeta_{I}(\eta)} \,\,\rmd\mu_{\rm L} } \nonumber \\
&=& \int\, e^{\f{i}{\sqrt{2}} \sum_{I} \bar\alpha_{I} \bar\zeta_{I}} \, e^{\f{i}{\sqrt{2}}\, \sum_{J}\alpha_{J} \zeta_{J}} \, \,\underbrace{\rmd \mu_{\rm B}} \nonumber\\
&=& \sum_{n=0}^{\infty} \sum_{m=1}^{\infty} \, (\f{i}{\sqrt{2}})^{m+n}\, \sum_{I=1}^{2}  \sum_{J=1}^{2}\, \f{\bar\alpha_{I}^{m} \alpha^{n}_{J}}{\sqrt{m!n!}} \,\, \int  \f{(\bar\zeta_{I})^{m}}{\sqrt{m!}}\, 
\f{(\zeta_{J})^{n}}{\sqrt{n!}}\,\, \rmd\mu_{\rm B} \nonumber\\
&=& e^{-\f{1}{2} \sum_{I} \alpha_{I}(\eta) \bar\alpha_{I}(\eta)} \, ,\ea
where in the third step we just expanded out the exponentials, and in the last step used the fact that 
the functions $\zeta^{n}_{I} /\sqrt{n!}$ constitute an orthonormal basis with respect to the Bargmann measure $\rmd\mu_{\rm B}$ (see Eq. (\ref{bargbasis})).

Thus, we again have the equality between the expectation values of the Weyl operators in the state $\Psi_{o}$ and those of their classical analogs in the state $\rho_{o}$:
\be  \label{key4}  \langle \hat{W}(\lambda,\,\mu)(\eta)  \rangle_{{}_{o}} =  \langle W(\lambda,\,\mu)(\eta) \rangle_{\rho_{o}} \ee
for \emph{all complex parameters $\lambda, \mu$, and for all times $\eta$}. By taking successive derivatives of this equality w.r.t. $\lambda, \bar\lambda, \, \mu, \bar\mu$ we obtain the equality of the VEVs of the Weyl ordered products of operators $\hvp_{I}(\eta)$ and $\hat\pi_{J}(\eta)$ with the expectation values of their classical counterparts:
\be \label{key5}\langle\big( F(\hvp_{I}(\eta), \hat\pi_{J}(\eta))\big)_{{}_{W}} \rangle_{{}_{o}} \,=\, \langle F\big(\vp_{I}(\eta), \pi_{J}(\eta)\big) \rangle_{\rho_{o}} \ee
for all polynomial classical observables $F(\vp_{I}, \pi_{J})$. This result provides a strong  justification for replacing the quantum state $\Psi_{o}$ with the classical state $\rho_{o}$. Note, in particular, that  the anti-commutator $ [\hvpk(\eta) ,\, \hat\pi_{\vk^{\prime}} (\eta)]_{+}$ we used in Section \ref{s2} is already Weyl ordered, whence its VEV  can be regarded as a classical quantity. This consideration provided a motivation for regarding  the ratio of the VEV of the commutator with that of the anti-commutator as a measure of `quantumness' of the dynamical phase under consideration. 

We will conclude with a number of remarks clarifying the implications and  limitations of this result.\\

\emph{Remarks:}

1. Let us begin with a mathematical detour. In the above analysis, for simplicity we restricted ourselves just to 2 modes $\vk$ and $-\vk$. However, the Bargmann representation exists rigorously also for the full quantum field with its infinite number of degrees of freedom \cite{bargmann2} (see also \cite{am-grg,segal}).  In particular, the measure $\rmd\mu_{\rm B}$ is rigorously defined, (although it can be split into a Gaussian pre-factor $e^{-\sum_{n} \zeta_{n} \bar\zeta_{n}}$ and a Liouville measure $\rmd\mu_{\rm L}$ only if the number of degrees of freedom is finite). The vacuum state is again fully characterized by the VEVs of  Weyl operators, and defines a classical mixed state canonically  which, however, has to be taken to be the Bargmann measure $\rmd\mu_{\rm B}$ itself; there is neither a distribution function nor the Liouville measure. Therefore, in contrast to Sections \ref{s3} and \ref{s4}, the classical behavior emerges in the sense of this section \emph{for the full quantum field}; one does not have to restrict oneself to a finite number of modes.

2.  Let us then consider an arbitrary number of modes $\vk_{1},\, \ldots\, \vk_{m}, \ldots\,$.  Now, if the phase space function $F$ contains only the configuration variables $\vp_{\vk}$ (or only the momentum variables $\pi_{\vk}$) then the Weyl ordering trivializes because all operators in the argument of $F$ commute.  Therefore from (\ref{key4}) we conclude 
\be  \langle\big( F(\hvp_{\vk_{1}}(\eta),\, \ldots\, \hvp_{\vk_{m}}(\eta),\,\ldots)\big) \rangle_{{}_{o}} \,=\, \langle \big( F(\vp_{\vk_{1}}(\eta), \ldots\, \vp_{\vk_{m}}(\eta),\,\ldots)\big) \rangle_{\rho_{o}}\, . \ee
(Similarly for a function only of the momentum variables $\hat\pi_{\vk_{1}},\, \ldots\,  \hat\pi_{\vk_{m}},\, \ldots$ \,.)  Thus, for any n-point function that involves only the configuration (or only the momentum) observables, we have an \emph{exact equality} between the classical and the quantum theory.  For these observables, we do not lose any information at all if we use the phase space distribution function $\rho_{o}$ in place of the quantum state $\Psi_{o}$. As noted in section \ref{s5.1}, mathematically  this equality is rather trivial. However,  it is of interest because currently one measures only these $n$-point functions of field operators. 

3. When would this procedure be inadequate? Let us set aside practical difficulties and assume for a moment that we would be able to measure arbitrary $n$-point functions involving both $\hvpk$ and $\hat\pi_{\vk}$ sometime in the distant future. If we were interested only in the Weyl ordered quantum operators, then no matter how large an $n$ we choose, the quantum prediction will again agree \emph{exactly} with the classical. So, the difference will be significant only if we were interested in measuring an $n$-point function that involves both the field operators and their conjugate momenta \emph{and} we have to perform a large number of permutations on these operators to bring the result to the Weyl-ordered form. This is a complete characterization of the limitation of the replacement of $\Psi_{o}$ with $\rho_{o}$.

4. Furthermore, these considerations \emph{are not tied to inflation}. Indeed in this section we worked on an arbitrary FLRW space-time. In fact, the results are vastly more general. We can consider quantum fields on any globally hyperbolic space-time. Given a complex structure $J$ (that is compatible with the symplectic structure $\Omega$) on the covariant phase space $\Gamma$, we can again construct the Bargmann representation. Given any foliation of the space-time by constant-time $t={\rm const}$ hypersurfaces, we can introduce configuration and momentum operators $\hvp(f) (t)$ and $\hat\pi(g)(t)$, smeared with test functions $f(\vx)$ and $g(\vx)$, and work with the algebra they generate.  In the Bargmann representation, the vacuum (defined by $J$)  is again represented by $\Psi_{o} =1$ and the Bargmann measure (constructed from the Riemannian metric $g$ on $\Gamma$ defined by $J$) again provides us with a mixed state, now represented as the Bargmann measure $\rmd\mu_{\rm B}$ on $\Gamma$.  And again one can show that the VEVs of \emph{any} Weyl ordered polynomials of $\hvp(f)$ and $\hat\pi(g)$ is \emph{exactly equal} to the expectation values of their classical analogs on this (mixed) state. 

5. Returning to cosmology, our results in sections \ref{s3} -\ref{s5} are directly applicable to tensor modes, and also to scalar modes in the inflationary scenario.  Therefore each provides a precise sense in which classical behavior emerges in the early universe, bringing out a few subtleties.  The generality of the main results of Section \ref{s5}, discussed above, may seem surprising at first. However, one should bear in mind the key assumption on which they rest: dynamics of perturbations is assumed to be linear. Therefore in the structure formation epoch where non-linearities play a dominant role, our considerations do not apply directly, (although  they can serve as a starting point in approaches that incorporate the effect of nonlinearities  perturbatively). On the one hand, this is a clear limitation of our approach. On the other hand, it shows that non-linear effects such as mode-mode coupling and decoherence --although important for independent reasons-- are not essential for the emergence of classical behavior in the early universe. 

\section{Discussion}
\label{s6}

As summarized in Section \ref{s1}, the issue of emergence of classical behavior has drawn a great deal of attention in the cosmology literature from different perspectives because of its conceptual as well as practical importance (see, in particular,  \cite{netoetal,guthpi,lpg,albretcht,dpaas,lps,kiefer1,kiefer2,psd,decoherence1,decoherence2,decoherence3,decoherence4,decoherence5,decoherence6,decoherence7,decoherence8,decoherence9,ls,decoherence10,jmvv1,sss,jmvv2,jmvv3,jmvv4}). 
In most of the leading scenarios, the origin of the observed anisotropies in the CMB  is traced back to vacuum fluctuations of quantum fields in the very early universe.  At the same time, there is a general expectation that the universe can be described using classical terms even in its early history. Indeed, in detailed treatments, one replaces the quantum vacuum by a distribution function on the classical phase space  early on --say, at the end of inflation--  and then describes the subsequent evolution entirely in classical terms.  Therefore several questions naturally arise: Is there something specific about the dynamics of cosmological perturbations, and/or the quantum state they are in, that naturally leads to the emergence of classical behavior a little later? Can we sharpen the sense in which this emergence occurs? Is it tied with inflation or is the phenomenon much more general? We addressed these issues in Sections \ref{s3} - \ref{s5}. Each of these sections focused on a  specific mechanism that has been used to argue why and how classical behavior can emerge in the early universe.  

Perhaps the most striking feature that distinguishes quantum mechanics from classical physics is the non-commutativity of observables. Therefore, if the quantum evolution of a system admits a phase in which `the importance of non-commutativity were to diminish', one could say that the system behaves classically in that phase. It has been argued that classicality  does emerge in this sense during inflation (see, e.g., \cite{dpaas,kiefer1,kiefer2} and references therein). However, as we discussed in Section \ref{s3} (and Appendix \ref{a1}), the specific reasoning used to arrive at this conclusion is often flawed. Specifically, the statement that the commutator between $\hvpk(\eta)$ and $\hat\pi_{\vk^{\prime}}(\eta)$ becomes negligible as a result of inflation is incorrect because, as is well known, the canonical commutation relations are preserved by quantum dynamics. Similarly, in Appendix \ref{a1} we show that the  commutator of field operators  ($\hat\phi_{\vk}(\eta_{1})$ and $\hat\phi_{\vk^{\prime}}(\eta_{2})$)  at different times also does not become negligible. Nonetheless, non-commutativity does fade during inflation in a certain precise sense:  While neither the commutators nor their VEVs  become negligible we showed that, as inflation unfolds, the VEV of the commutator becomes negligible relative to the VEV of the anti-commutator (both, for the canonically conjugate operators, and for field operators at different times).  For the canonically conjugate operators, the ratio of the VEVs of the commutator and the anti-commutator decreases exponentially with the number  of e-folds. For field operators at two different times, we found  that the behavior of the ratio is much more subtle. In particular, the ratio at a later time $\eta_{2}$ is suppressed relative to that at an earlier time $\eta_{1}$ only if $\eta_{1}$ is taken to be a few e-folds after the mode exits the Hubble horizon. 

In Section \ref{s4} we discussed another notion that has been taken to be a hallmark of the emergence of classical behavior: the phenomenon associated with quantum squeezing (see, e.g., \cite{lpg,albretcht,dpaas,lps,kiefer1,kiefer2,jmvv1}). Again, most of the cosmological discussion we are aware of is in the context of inflation, where at late times the uncertainty in the field variable $\hvpk(\eta)$ continues to be squeezed (approaching half its value in the vacuum state of the field in Minkowski space-time).  The uncertainty in $\hat\pi_{\vk}(\eta)$ keeps growing exponentially in the number of e-folds as time evolves, but this growth does not contaminate the sharply peaked nature of the state in the variable $\hvpk(\eta)$ as time passes.  We provided a simple physical explanation for this phenomenon using intuition from quantum mechanics of macroscopic systems (such as a grandfather clock). 

More importantly, we traced the phenomenon of squeezing back to the classical evolution on the \emph{phase space} $\Gamma$. This was possible because one can succinctly isolate the new mathematical structure that is needed in the passage from the classical to the quantum theory of fields under consideration: One has to supplement the classically available symplectic structure $\Omega$ with a Riemannian metric $g$ on $\Gamma$ (such that the pair $(g,\Omega)$ equips $\Gamma$ with a K\"ahler structure \cite{am,am-grg}).  While the symplectic inner-product between canonically conjugate vectors associated with $(\vpk,\,\pi_{\vk})$ is preserved under time-evolution, their metric inner product is not. As time evolves, the norm of one of these vectors is squeezed while  that of the other is stretched. And this behavior is faithfully mirrored in the squeezing and stretching of quantum uncertainties.  This  classical characterization of squeezing illuminates some of the finer aspects of the phenomenon. For example, one finds that during inflation almost all of the  squeezing occurs before the mode $\vk$ exits the Hubble horizon; after the exit there is only a factor of 2 decrease in the norm of the Hamiltonian vector field even if one waits forever. Furthermore, one finds that inflation plays no essential role in this discussion. In the radiation or dust dominated universe, for example, not only does this classical squeezing occur but is even more pronounced because the norm of the Hamiltonian vector field generated by $\vpk$ continues to decrease forever; in de Sitter, it reaches a non-zero asymptotic limit.

In calculations that bridge the quantum field theory of cosmological perturbations with observations, one replaces the quantum state $\Psi_{o}$ of the field by a mixed classical state on the phase space $\Gamma$, and the quantum evolution by the classical Hamiltonian flow on $\Gamma$.  Off hand, the procedure may seem ad-hoc. But if the answers provided by the full quantum theory are well approximated, not only would the procedure be justified but it would also provide a clear-cut demonstration of the emergence of classical behavior. In Section \ref{s5} we analyzed this issue using the geometrical structures on $\Gamma$ introduced in Section \ref{s2}. These structures naturally lead one to the Bargmann representation of the quantum algebra in which states $\Psi$ are represented by (holomorphic) functions \emph{on the phase space} $\Gamma$.  Thus,  unlike in most discussions of this issue, quantum states are not represented by functions just of $\vpk$ or just of $\pi_{\vk}$; the phase space symmetry between $\vpk$ and $\pi_{\vk}$ is not broken, making it much easier to locate the desired correspondence between quantum and classical behavior. In particular, the classical mixed state defined by the quantum vacuum $\Psi_{o}$  is just the Bargmann measure $\rmd \mu_{\rm B}$ on the infinite dimensional phase space $\Gamma$. Furthermore, this correspondence leads to a simple and remarkably general relation between quantum and classical $n$-point functions:
\ba \label{key6}  &\langle&\big( F(\hvp_{\vk_{1}}(\eta),\, \ldots\, \hvp_{\vk_{m}}(\eta),\,\ldots  \hat\pi_{\vk_{1}}(\eta)  \ldots \hat\pi_{\vk_{m^{\prime}}}(\eta) )\big)_{{}_{}W} \rangle_{{}_{o}} \nonumber\\
&=&\, \langle \big( F(\vp_{\vk_{1}}(\eta), \ldots\, \vp_{\vk_{m}}(\eta),\ldots \, \pi_{\vk_{1}}(\eta)   \ldots \pi_{\vk_{m^{\prime}}}(\eta)  \ldots)\big) \rangle_{\rmd \mu_{{\rm B}}}\,  \ea
%
%
for \emph{any} polynomial $F$ and any time $\eta$,  where the suffix $W$ stands for `Weyl ordering'.  Thus, (essentially) every classical observable ${\mathcal{O}}$ admits an explicit quantum analog  $\hat{\mathcal{O}}_{\rm W}$ whose VEV exactly equals the expectation value of ${\mathcal{O}}$ in the classical probability distribution measure $\rmd \mu_{\rm B}$  \emph{for all times} $\eta$. This equality brings out the precise --and surprisingly strong-- sense in which $\Psi_{o}$ and quantum dynamics can be replaced by $\rho_{o}$ and classical dynamics. 
Currently, one observes only the 2- (and 3-) point functions involving just the field variable $\hvpk(\eta)$. For these observables, the Weyl ordering trivializes since  $\hvpk(\eta)$ and $\hvp_{\vk^{\prime}}(\eta)$ commute and one recovers the result that has been obtained by other methods \cite{dpaas,lps,kiefer1,kiefer2,jmvv1}.

Thus the three sections explored three different senses in which classical behavior can emerge in the early universe. All three are realized in inflationary scenarios. Since most discussions of the issue of emergence of classicality have been in the context of inflation, there appears to be a general impression that these are different facets of the same underlying notion. However, we saw that this is not the case: Emergence in one sense does not imply emergence in another. In the radiation or dust filled universes --or those that interpolate between the two in the sense of Section \ref{s2.2}-- classicality does emerge in the sense of Sections \ref{s4} and \ref{s5} but not in the sense of Section \ref{s3} or Appendix \ref{a1}. Similarly, there are systems (such as the grandfather clock) which clearly behave classically in the intuitive physical sense but in which only the criterion of Section \ref{s5} is met. We also saw that at a conceptual level inflation is not essential for semi-classicality to arise in the early universe in the sense of Sections \ref{s4} and \ref{s5}. However, because there is a large number of e-folds \emph{in a very small interval of proper time}, squeezing, for example, occurs \emph{extremely}  fast in proper time during inflation. 

Finally, the last notion of classical behavior discussed in Section \ref{s5} is very general. In particular, one \emph{does not have to restrict oneself to a finite number of modes;} the Bargmann representation exists rigorously for the full field with its infinite number of degrees of freedom \cite{bargmann2,segal} and the correspondence (\ref{key6}) between the quantum vacuum and the classical mixed state holds for all quasi-free vacua (as well as for all coherent states in the Fock spaces defined by these states). To our knowledge, this exact correspondence has not been discussed in the early universe literature.%
\footnote{See, for example,  Ref. \cite{jmvv1} in the cosmological literature where it is shown that the equality between quantum and classical expectation values holds, but only  ``as far as two-point correlators are concerned'' (discussion between Eq.(104) and the end of Section V in the version arXiv:1510.04038v7 [astro-ph.CO] 11 Jul 2019).  But after our pre-print appeared in the arXiv, Lajos Di\'osi  informed us that the analog of (\ref{key6}) does appear in (Eq. (17) of) Ref. \cite{diosi4}. However, since that article is on weak measurements in non-relativistic quantum mechanics, it does not discuss quantum field theoretic issues associated with an infinite number of degrees of freedom, cosmological perturbations, and the Bargmann representation.}
In particular, as far as we are aware of, in the Wigner function approach to the relation between quantum and classical predictions all discussions are restricted to finite dimensional phase spaces since the underlying Liouville measure $\rmd \mu_{\rm L}$ does not admit an extension to infinite dimensions. Therefore, in the cosmological context,  discussions based on the Wigner functions are restricted  to a finite number of modes. (In the literature, one often considers just two, labelled by $\vk$ and $-\vk$).  By contrast in our approach, as we just saw,  the classical (mixed) state corresponding to any quasi-free vacuum is the Bargrmann measure $\rmd \mu_{\rm B}$ -- on the infinite dimensional phase space. It is only if one restricts oneself to a finite number of modes that the measure can be split as $\rmd \mu_{\rm B} = \rho_{o} \rmd \mu_{\rm L}$, and one can represent the classical state as a distribution \emph{function} $\rho_{o}$ as in the Wigner approach. Another important aspect of generality is that the discussion is not tied to the cosmological setting but can be extended to quantum fields  on any globally hyperbolic space-time, so long as they satisfy a linear field equation.

Our goal was to analyze conceptual issues related to emergence of classicality in the early universe from the perspective of mathematical physics. Therefore we focused on the \emph{simplest context} in which these questions can be  answered and details of why and how classical behavior emerges can be worked out.  This is why we emphasized  `emergence' rather than `quantum to classical transition', and also why we stayed away from non-linearities in dynamics, the issue of mode-couplings, environment degrees of freedom and decoherence, the quantum discord and the quantum measurement theory. These issues are clearly important and have to be addressed in more complete descriptions. But, as many in the community have argued, the issues of interest here can be addressed without having to include these additional considerations. Our discussion brought out the precise sense in which classical behavior emerges in the early universe already in the simplest context --that of the quantum theory of linear cosmological perturbations.%
\footnote{If we were to use an analogy with  the physics of the hydrogen atom, our focus would be analogous to calculating the energy levels, degeneracies, and understanding the origin of degeneracies in spectral lines, rather than on how the hydrogen atom makes a transition from one level to another, the mechanism by which a the photon is emitted, and whether the wave function collapses during this measurement process.}

\section{Note Added in response to arXiv:2009.09999}
\label{s6}

Javier Berjon, Elias Okon, and Daniel Sudarsky (referred to as BOS in what follows)  submitted an article to a journal. Following their general policy, Editors asked us to comment on this paper 
because  it ``appears to be critical of some aspects of reference 1 of the manuscript, which you coauthored'' (namely, this paper).   We responded to the journal with the following comments. Since the arXiv submission has remained unchanged, we are now putting our response in the public domain. \\

1. We are perplexed and surprised by this paper. Perplexed, because while the wording in BOS suggests that our paper ``Emergence of classical behavior in the early Universe'', Physical Review D 102, 023512 (2020) (referred to as ACK below) is the center of their criticism,  the critical statements refer to claims in other, older works, not to results in our paper. 
Surprised, because while already in their abstract the BOS paper criticizes the literature for ``lack of clarity'' because of reliance on ``unjustified and implicit assumptions'', we encountered these very problems in the BOS paper! In what follows we  explain these two  points in some detail.\\

2. The abstract of the ACK paper emphasizes that we \emph{investigate} three issues that have been discussed in the context of inflation, and the concluding para of the paper emphasizes that the goal was to discuss these issues \emph{from a mathematical physics perspective.}  We also emphasize in this para that issues such as non-linearities, mode-mode couplings, decoherence, quantum discord and quantum measurement theory \emph{are important  and will have to be addressed in a more complete discussion}.  In contradistinction to what the wording used by BOS suggests,  we never say that our results provide a complete account.  Nowhere did we claim or suggest that any of the three criteria we discuss implies that the state becomes inhomogeneous and anisotropic, or leads to what BOS refer to as ``classicalization''. In fact, already in our Introduction,  we emphasize that in our analysis ``there is no quantum to classical transition'', whence we ``will not need to enter a discussion of issues that arise when the focus is on transition''.  We also say ``Rather, our emphasis is on the ``emergence'' of classical behavior ... in the mathematical description of cosmological perturbations'' and our notion of emergence refers to the dynamics of specific sub-classes of observables. Therefore, the main thrust of the negative comments in the BOS paper has very little to do with what we actually discuss and prove. \\

3.  What we actually do in ACK is to carefully examine the three criteria that have been widely used in the literature in the discussion of the emergence of semi-classical behavior and clarify the precise sense in which the statements in the literature hold  --sometimes upon appropriate modifications-- and the sense in which some of the mathematical arguments are incomplete, or even incorrect.  So the emphasis is on  a \emph{critical mathematical investigation}. In particular, some of the literature had implicitly  assumed that the three criteria are different facets of the same underlying phenomenon and we show that this is not the case.  

It is only the Section 4  of the BOS paper that directly addresses the results of ACK.  
(i) They agree with us  that the first criterion --fading of non-commutativity-- has some important limitations.\\ 
(ii) The BOS summary of what motivated previous works to use squeezing as a signature of the emergence of classical behavior is the same as ours.  But then they say that in the ACK  paper there  are ``no actual arguments presented in favor of quantum squeezing as a sign of the emergence of classical behavior''.  This is incorrect.  ACK have emphasized that their use of the term ``classical behavior'' refers to a set of observables and their uncertainties in a given state. (BOS also state this as their own viewpoint on page 8).  Classical behavior emerges for the field operators in which the uncertainty is squeezed, and ACK also provide a physical understanding as to why it continues to remain squeezed as time evolves in spite of the fact that the uncertainty in field momenta is large. The main point of this Section in ACK is to trace back the phenomenon of  squeezing to geometrical structures on the classical phase space. BOS have no comment on that.\\
(iii) The third criterion is perhaps the most important one ``in practice'' because as ACK explain --and BOS reiterate-- in most calculations one generally replaces the quantum state of perturbations by a distribution function on their classical phase space. Because a state is completely determined by the expectation values of all observables both in classical and quantum mechanics,  \emph{time evolutions} of expectation values provide a natural avenue to critically investigate the validity of this procedure. Using the Bargmann representation for quantum states of perturbations, we obtained a sharp mathematical result on when the procedure is justified, and when it is not. Our results are more general than what was known in the cosmology literature before. We find that the class of observables is surprisingly large, much larger than products of just the field operators, or, just their momenta, that were typically considered. For the larger class of observables, then, the mathematical procedure used in the literature to time-evolve these observables is justified. The BOS paper has no comment on this main result. Their criticism is that this procedure does not explain the breakdown of homogeneity and isotropy.  ACK never said that it does. \\
Thus, the criticisms in the BOS Sections 4 and 5 have almost nothing to do with the ACK results.
\\

4. In the BOS paper, we found several instances of ``lack of clarity'' because of reliance on ``unjustified and implicit assumptions''. 

Page 6:  ``It is important to point out that, in order to apply our proposed criterion for classicality, it is not necessary for actual measurements to ever take place; that is, a system can be deemed to have classicalized, even if there are no observers around to verify it.  Still the notion of measurements remains central, as it does in the standard quantum interpretation.''\\
Central, for what? Presumably not for ``classicalization'', since that is what the previous sentence states. But isn't the entire BOS focus on what they call ``classicalization''? 

Page 7: ``However, it becomes untenable as soon as one intends to employ the theory in more complex scenarios, such as the early universe, in which there are no observers around to measure, ...''\\
But did't BOS say on page 6 that criterion for ``classicality'' does not need an observer? 

Page 8: ``In particular, we saw that classicalization can only be defined relative
to a set of observables, with associated uncertainties, and that a given state of the
system might be describable in classical terms when focusing on a certain feature, but
not when focusing on others.''\\
This is a reiteration of the ACK viewpoint. But then we are puzzled by the ``problem''  discussed on page 18:\\
``The problem, of course, is that, just because a certain variable has become squeezed, does not entail that other variables will too.''\\
 Why is this a problem? After all,  criterion BOS fully accept on page 6,  ``there is no absolute sense in which a quantum system might be said to classicalize this can only be asserted relative to the cited additional information.''

Page 9: ``Now, in the same way that a breakdown of homogeneity and isotropy does not
imply classicality, classicality can occur in the absence of a breakdown of the symmetry.
Nevertheless, what we want to point out is that, classicalization by itself, cannot erase
or eliminate a symmetry present at the quantum level and, in particular, it cannot
break homogeneity and isotropy.''\\
ACK never say that emergence of the classical behavior in any of the senses discussed implies breaking of homogeneity and isotropy. So what is the point that BOS are making?

Page 10: ``any successful effort to explain the emergence of classical behavior in the standard cosmological context must necessarily involve a satisfactory account of the breakdown of the homogeneity and isotropy of the quantum state.''\\
This seems to contradict their assertion on page 9 that the two are unrelated. Also,  this conclusion seems to be too sweeping and therefore lacking in precision/clarity.  In fact the quote above from page 8 says that this ``emergence'' can be defined relative to a set of observables and that procedure does not require breakdown of homogeneity and isotropy.\\

5. \emph{Summary:} The only way we can possibly understand the perplexing BOS submission is through the concern they express in the opening para of their Section [4]: The ACK work could ``then be interpreted as an attempt to overcome'' the criticisms of  [8,10,19,24]  ``by strengthening the standard arguments'' made, e.g. in [2, 6,12-16,18, 23].%
\footnote{Refs.  [8,10,19, 24] refer to older works in inflation, not to what we actually do (see Point 3 above).  [10] is on the Continuous Spontaneous Localization idea and [19] is a criticism of that idea;  they are not even tangentially related to our results! [8, 24] are co-authored by Sudarsky.}
In light of what we have said above, this is clearly a misinterpretation made by BOS. Their entire paper appears to be a reaction in defense of the arguments made in [8,24], based on this misinterpretation.

\begin{appendix}

\section{Fading of non-commutativity between $\hphik(\eta_{1})$ and $\hat\phi_{\vec{k}^{\prime}}(\eta_{2})$ } 
\label{a1}
 
Since the field operators $\hphik(\eta)$ and $\hat\phi_{\vec{k}^{\prime}}(\eta)$ at the same time $\eta$ commute, we need to consider the commutator and the anti-commutator between field operators evaluated at two different times $\eta_{1}$ and $\eta_{2}$.  Now, it is often said that  
``for modes which presently appear on large cosmological scales, the ratio of growing mode to the decaying mode is  $\propto 10^{-100}$\,'' and consequently, ``one also has [$\hphik(\eta_{1}),\,\hat\phi_{\vec{k}^{\prime}}(\eta_{2})] \approx 0$'' (using our notation for field operators).  As a result, ``if a measurement puts the system into an eigenstate of $\hphik(\eta_{1})$, all future measurements would give the corresponding eigenstates of $\hphik (\eta_{2})$ for $\eta_{2} > \eta_{1}$,  corresponding to the classical evolution of the system'' (see, e.g. \cite{kiefer2}). In this Appendix we sharpen  the sense in which these considerations hold. As in section \ref{s3},  in the main discussion we will use ratios of the expectation values of commutators and anti-commutators, and then comment on the commutators themselves.
 
Let us first calculate the vacuum expectation value of the product $\hphik(\eta_{1})\, \hphik(\eta_{2})$, where $\eta_{2}$ is to the future of $\eta_{1}$.  Using the explicit form (\ref{phipi}) of field operators and (\ref{BD}) of basis functions, we obtain:
\be 
\langle\,\,\hphik (\eta_{1})\, \hat{\phi}_{\vk^{\prime}} (\eta_{2})\,\, \rangle=   \hbar\,  e_{k}(\eta_{1}) \bar{e}_{k}(\eta_{2})\,   \delta_{\vk ,\, -\vkp} \, .\ee
Therefore the absolute value of the ratio of the expectation value of the commutator to that of the anti-commutator is of interest only if $\vec{k} = - \vec{k}^{\prime}$ and is then given by 
\be \label{ratio1}
|\,R_{\hphik(\eta_{1}),\,\hat{\phi}_{-\vk}(\eta_{2})} \,| = \Big{|} \f{ \mathfrak{Im}\, e_{k}(\eta_{1}) \bar{e}_{k}(\eta_{2})}{\mathfrak{Re}\, e_{k}(\eta_{1}) \bar{e}_{k}(\eta_{2})} \Big{|}. \ee
Using the expression (\ref{BD}) of $e_{k}(\eta)$ in de Sitter space-time, one obtains 
\be \label{product} e_{k}(\eta_{1}) \bar{e}_{k}(\eta_{2}) = \Big(\f{1}{a(\eta_{1})} + i \f{H}{k}\Big)\, \Big(\f{1}{a(\eta_{2})} - i \f{H}{k}\Big) \, 
\f{e^{ik(\eta_{2}-\eta_{1})} }{2k}.\ee
Since the significance of non-commutativity is expected to fade after the mode exits the Hubble horizon, let us introduce three instants of conformal time: $\eta_{k} = -1/k$, the mode exit time;  and $\eta_{1}, \eta_{2}$, such that $\eta_{2}$ is to the future of $\eta_{1}$ which is to the future of $\eta_{k}$.  Let us suppose there are $N_{0}$ e-folds between $\eta_{k}$ and $\eta_{1}$ and $N$ e-folds between $\eta_{1}$ and $\eta_{2}$, so that 
\be \f{a(\eta_{1})}{a(\eta_{k})} = e^{N_{0}} \qquad {\rm and} \qquad  \f{a(\eta_{2})}{a(\eta_{1})} = e^{N}\, . \ee
Then, using (\ref{product}) and the fact that $a(\eta_{k}) = \f{k}{H}$,  one obtains:
\be \label{ratio2} |\, R_{\hphik(\eta_{1}),\,\hat{\phi}_{-\vk}(\eta_{2})}\, | = \f{(1 + e^{- 2 N_0 - N}) \sin[e^{-N_0}(1 - e^{-N})] - e^{-N_0}(1 - e^{-N}) \cos[e^{-N_0}(1 - e^{-N})]}{(1 + e^{- 2 N_0 - N}) \cos[e^{-N_0}(1 - e^{-N})] + e^{-N_0}(1 - e^{-N}) \sin[e^{-N_0}(1 - e^{-N})]} \, .  \ee
(Note that if N=0, i.e., $\eta_{1} = \eta_{2}$, the ratio vanishes as it must since the field operators commute in this case.) \\ 

Eq. (\ref{ratio2}) has three interesting implications:
\begin{itemize}
\item We are interested in the case in which $\eta_{2}$ is sufficiently to the future of $\eta_{1}$ so that $e^{-N} \ll 1$.  The limiting behavior of  the ratio is given by
\be \label{ratio3}  \lim_{N\to \infty}\, \, |R_{\hphik(\eta_{1}),\,\hat{\phi}_{-\vk}(\eta_{2})} \, | \, = \,\,\f{\sin e^{-N_0} - e^{-N_0} \cos e^{-N_0}}{ \cos e^{-N_0} + e^{-N_0} \sin e^{-N_0}}   \ee
Because of the $e^{-N_{0}}$ factors, this limit  is reached \emph{extremely} rapidly.  For $N_{0} =0$, i.e., when $\eta_{1}$ is chosen to coincide with the horizon exit time for the mode,  the ratio differs from its asymptotic value $N\to \infty$ just by a $0.4\%$ already when $\eta_{2}$ is  just 2 e-folds after $\eta_{1}$. (See the left panel of  Fig. \ref{fig1}.)  For $N_{0} =2$, i.e., if $\eta_{1}$ is chosen to be 2 e-folds  after horizon crossing, then the approach to the asymptotic value $N\to \infty$ is even faster.  The difference is only $0.25\%$, again if $\eta_{2}$ is chosen to be 2 e-folds after  $\eta_{1}$. (See the middle panel of Fig. \ref{fig1}.)

\item However, Eq. (\ref{ratio3}) also brings out the a striking feature of dynamics: If $N_{0}=0$, i.e., $\eta_{1}$ is chosen to be the horizon crossing time $\eta_{k}$  for the mode, then the ratio (and hence the commutator) does \emph{not} vanish no matter how long we wait:
\be \label{limit}  \lim_{N\to \infty}\, |R_{\hphik(\eta_{k}),\, \hat{\phi}_{-\vk}(\eta_{2})} \, | \, = \,  
\f{\sin 1 - \cos 1}{\sin 1 + \cos 1} \,\simeq \,  0.218\, .  \ee 
Thus, the commutator of the field operator at time $\eta_{k}$ and at any later time $\eta_{2}$ remains $O(1)$, no matter how many e-folds $N$ one waits.  From a general conceptual viewpoint, this limit is interesting because it makes no reference to the mode, or indeed to $H$, i.e., the value of the cosmological constant of the (Poincar\'e patch of the) de Sitter background.  For \emph{any} mode $k$, the value of the ratio between the time $(\eta= -(1/k))$ that mode exits the Hubble horizon and asymptotic future ($\eta=0$) is a fixed number given by the right side of (\ref{limit}). This number is thus an interesting invariant of the quantum field theory under consideration.

\begin{figure}[]
  \begin{center}
  \vskip-0.4cm
   \includegraphics[width=1.8in,height=2in,angle=0]{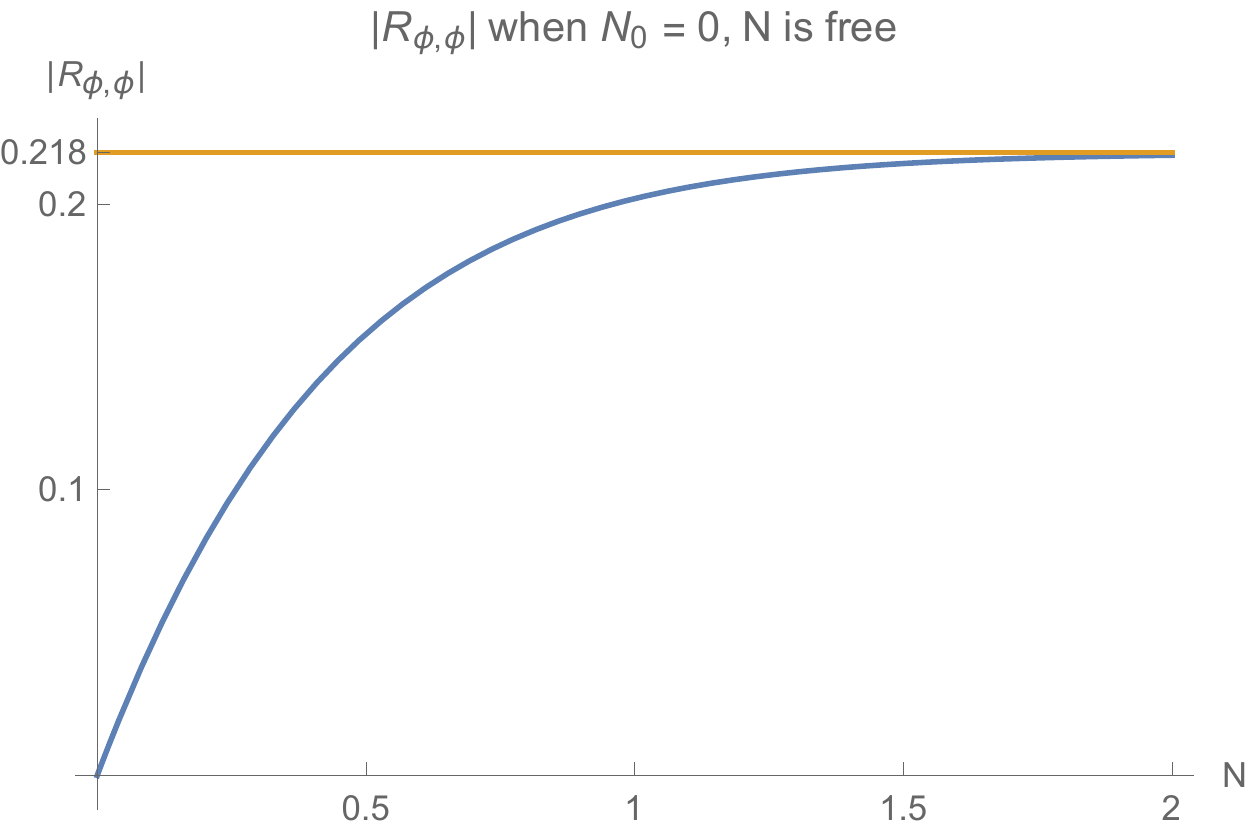}\hskip1cm
   \includegraphics[width=1.8in,height=2in,angle=0]{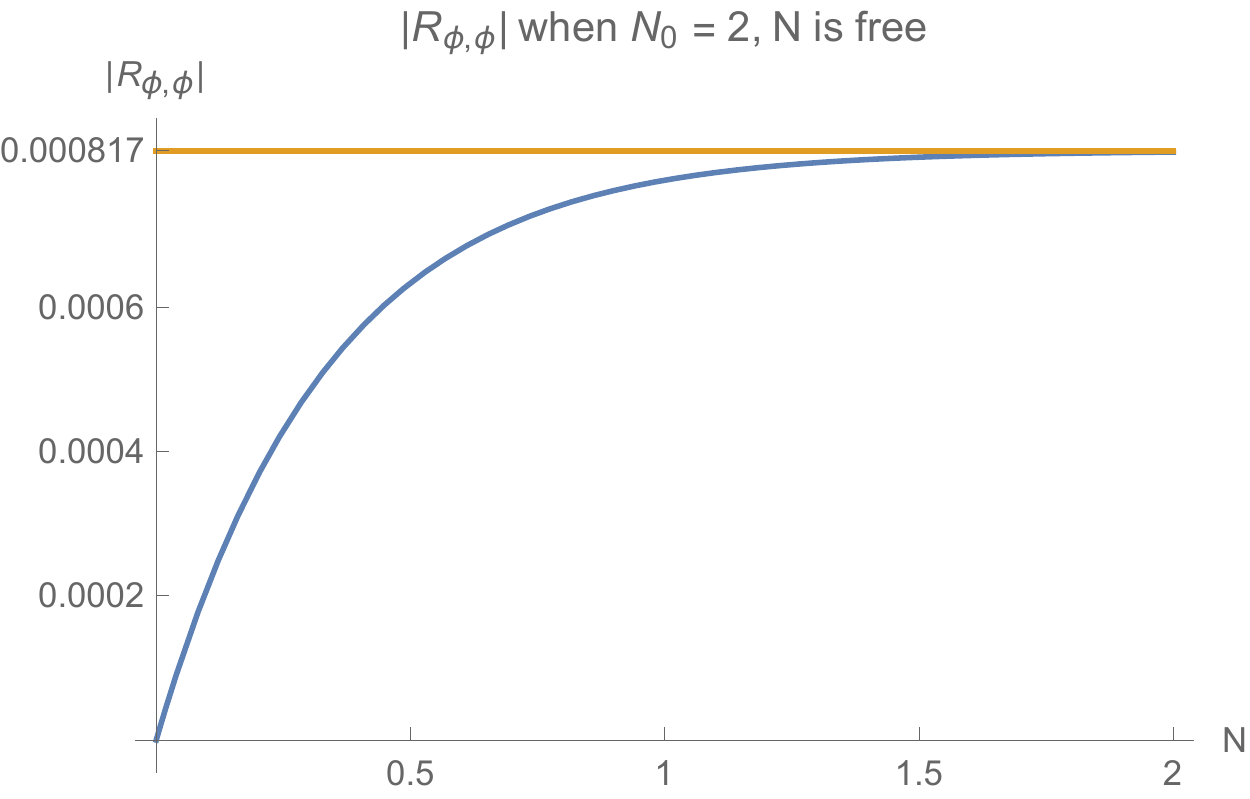}\hskip1cm
   \includegraphics[width=1.8in,height=2in,angle=0]{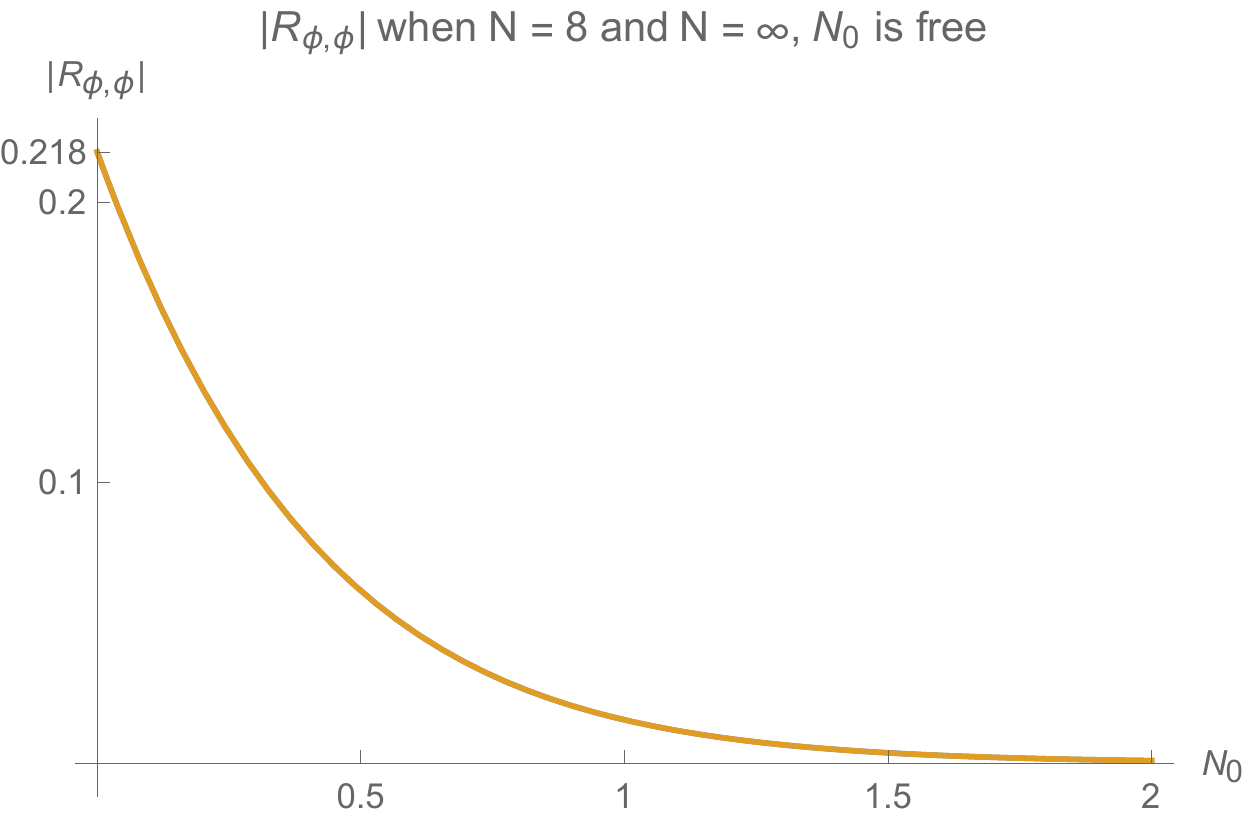}
\caption{The absolute value of  $R_{\hphik(\eta_{1}),\,\hat{\phi}_{-\vk}(\eta_{2})}$. 
$N_{0}$ is the number of e-folds between the horizon crossing time $\eta_{k}$ of the mode and $\eta_{1}$, while $N$ is the number of e-folds between $\eta_{1}$ and $\eta_{2}$. 
\emph{Left panel}:  $\eta_{1}$ is chosen to be the horizon exit time for the mode, so  $N_0 = 0$, while $N$ varies. The asymptotic value ($N\to \infty$) of the ratio is approached rapidly. Already at  $N = 2$, the ratio is differs from its asymptotic value by only $8.56 \times 10^{-4}$ or by $0.4 \%$.
\emph{Middle panel}: $\eta_{1}$ is now chosen to be 2 e-folds to the future of $\eta_{k}$, so $N_{0} =2$ and $N$ is again allowed to vary. Now the asymptotic value is reached even more rapidly.  For $N=2$, now the difference from the asymptotic value is reduced to $0.25\% $.
\emph{Right Panel:} Now $N$ is kept fixed and $N_{0}$ varies. The plots for $N=8$ and $N=\infty$ are indistinguishable at this resolution.}
\label{fig1}
\end{center}
\end{figure}

\item Now, if we set  $N_{0} =2$, i.e. if $\eta_{1}$ is taken to be just 2 e-folds after the mode exits the Hubble horizon, we find 
\be \lim_{N\to\infty}  |\, R_{\hphik(\eta_{1}),\,\hphi_{-\vk}(\eta_{2})}\, | \,\approx \,\   8.17 \times 10^{-4}\ee
and this number further decreases very rapidly as $N_{0}$ increases, i.e.,  if we choose $\eta_{1}$ to be a few more e-folds after the horizon crossing time $\eta_{k}$.  Again, the asymptotic value is reached very quickly for any choice of $N_{0}$ (i.e. for any $\eta_{1}$ to the future of $\eta_{k}$). The right panel of  Fig. \ref{fig1}  shows that the value of the ratio for $N=8$ is already essentially indistinguishable from the asymptotic value $N=\infty$ for any $N_{0}$.
\end{itemize}

Let us summarize. The vacuum expectation value of the commutator of the field operator at two different times can be obtained from the imaginary part of Eq. (\ref{product}),
\be \langle\,\,[\hphik (\eta_{1})\, , \hat{\phi}_{-\vk} (\eta_{2})]\,\, \rangle = 2 i \hbar\, \, \mathfrak{Im} (e_{k}(\eta_{1}) \bar{e}_{k}(\eta_{2}) )\,  , \ee
for any $\eta_{1}, \eta_{2}$. {  However, since this quantity dimensionfull, a priori we cannot say if it is small or large.}
%
%
If one works with the ratio $|R_{\hphik(\eta_{1}),\,\hat{\phi}_{-\vk}(\eta_{2})} |$ we obtain a dimensionless quantity and we can therefore ask if it is negligible compared to $1$. It is, provided we choose $\eta_{1}$ to be a few e-folds (as few as 2) after the horizon exit time, and $\eta_{2}$ to be say 8 or more e-folds after $\eta_{1}$. Thus,  even with ratios, the sense in which the importance of non-commutativity between unequal time field operators diminishes  is more subtle than that for the equal time canonically conjugate operators discussed in Section \ref{s3.1}.\\

\emph{Remarks:}\\

1.  Let us return to the commutator itself. The expression (\ref{phiop}) of the field operator implies that the commutator is given by the imaginary part of Eq. (\ref{product}),
\be \,[\hphik (\eta_{1})\, , \hat{\phi}_{-\vk} (\eta_{2})]\,\,  =  2 i \hbar\, \, \mathfrak{Im} (e_{k}(\eta_{1}) \bar{e}_{k}(\eta_{2}) )\, \hat{\mathbb{I}} , \ee
for any $\eta_{1}, \eta_{2}$.  { Since the right side is a multiple of identity, its expectation value independent of the state, and since -unlike in the canonical commutation relations considered in Section \ref{s3}-- it is time \emph{dependent},  we can analyze whether it fades in time by comparing its values for different pairs of times.} A natural strategy is to fix $\mathring\eta_{1}$ and $\mathring\eta_{2}$ (which is to the future of $\mathring{\eta}_{1}$) and ask if the \emph{dimensionless} ratio 
\be \label{ratio4} \f{\mathfrak{Im} (e_{k}(\mathring{\eta}_{1}) {\bar{e}}_{k} (\eta_{2}))}{\mathfrak{Im} (e_{k}(\mathring{\eta}_{1}) \bar{e}_{k}(\mathring{\eta}_{2}) )} \ee
goes to zero as we move $\eta_{2}$ to the future of $\mathring{\eta}_{2}$ for any given $k$. 
 
Since the general expectation is that the commutator would become negligible if $\eta_{2}$ is chosen to be in the future of the horizon crossing time, let us  set $\mathring\eta_{1} = \eta_{k}$  (so $N_{0} =0$), and $\eta_{2}$ to be in the asymptotic future (so $N\to \infty$), then 
\be 2 i \hbar\, \, \mathfrak{Im} (e_{k}(\mathring{\eta}_{1}) \bar{e}_{k}(\eta_{2}) )\, \hat{\mathbb{I}} 
  \to  i\, (\sin 1 - \cos 1)\,\f{H^{2}}{k^{3}}\,  \hbar\,\, \approx\, 0.30  i \,\,\f{H^{2}}{k^{3}}  \ee
{ which is non-zero, whence the dimensionless ratio (\ref{ratio4}) does \emph{not} go to zero irrespective of how we choose $\mathring{\eta}_{2}$.
In fact} if we were to choose $\mathring{\eta}_{2}$ to be a few e-folds $\mathring{N}$  after $\mathring{\eta}_{1}$ such that $e^{-\mathring{N}} \ll1$, then the ratio (\ref{ratio4}) is approximately 1.  Thus although, in contrast to the canonical commutation relations we examined in Section \ref{s3}, the unequal time commutators are time dependent, we do not see a precise sense in which non-commutativity fades as inflation proceeds.

2.  By contrast, the strategy of using ratios of expectation values of commutators and anti-commutators does provide a sense in which the significance of non-commutativity fades during inflation. The sense is direct for canonical commutation relations of section \ref{s3} and more subtle for  unequal time commutators discussed above.  However,  as we remarked at the end of Section \ref{s3}, examination of this strategy in more general contexts beyond inflation shows that this is not a robust signal of emergence of classical behavior.  Consider for definiteness a radiation-filled FLRW universe. In this case,  as we noted in Section \ref{s2}, one can exploit the fact that the scalar curvature vanishes and introduce a natural vacuum state.  Eq. (\ref{ratio1}) implies that the ratio of expectation values in this vacuum is given by
\be \big{|}\, R_{\hphik(\eta_{1}),\,\hat{\phi}_{-\vk}(\eta_{2})}\, \big{|}\, =\, \Big{|} \f{\sin k(\eta_{2} - \eta_{1})}{\cos(\eta_{2} -\eta_{1})} \Big{|} \, ,\ee 
for any $\eta_{1}$ and $\eta_{2}$. Thus, if we keep $\eta_{1}$ fixed and increase $\eta_{2}$, the ratio  simply oscillates between $0$ and $\infty$ (exactly as in Minkowski space-time). On the other hand, as discussed in Section \ref{s4} that in this case, expectation values of the canonically conjugate operators do exhibit the squeezing behavior  --the uncertainty in the field $\hvp_{\vk}(\eta)$ decreases and that in $\hpi_{\vk}(\eta)$ decreases as the universe expands.  So classical behavior does emerge in the sense that the state remains sharply peaked on the field variable as time evolves, { even though the non-commutativity does not fade.} 
\end{appendix}

\section*{Acknowledgments}
AA thanks Jerome Martin for a discussion and Lajos Di\'osi for correspondence. This work was supported by  the NSF grants PHY-1505411 and PHY-1806356 and the Eberly Chair funds of Penn State;  DGAPA-UNAM IN114620 and CONACyT 0177840 grants, and an Edward A. and Rosemary A. Mebus Graduate Fellowship in Physics and the Frymoyer Honors Fellowship at Penn State, to AK.

\end{document}